\documentclass[aps,pre,twocolumn,superscriptaddress]{revtex4-2}
\usepackage[utf8]{inputenc}
\usepackage{bm}
\usepackage{amssymb}
\usepackage{mathtools}
\usepackage{amsmath}
\usepackage{xcolor}
\usepackage{enumitem}
\usepackage{natbib}
\usepackage{hyperref}
\usepackage{lipsum}
\usepackage{lineno}
\usepackage[ruled,vlined]{algorithm2e}

\begin{document}
%
\title{Including the magnitude variability of a signal into the ordinal
pattern analysis}
\author{Melvyn Tyloo}
\affiliation{University of Exeter, Living Systems Institute, EX4 4QD Exeter, United Kingdom}
\affiliation{University of Exeter, Department of Mathematics and Statistics, EX4 4QD Exeter,
United Kingdom}
\author{Joaqu{\'i}n Gonz{\'a}lez}
\affiliation{New York University Grossman School of Medicine, Neuroscience Institute and Department of Psychiatry, New York, USA}
\author{Nicol\'as Rubido} \email{nicolas.rubidoobrer@abdn.ac.uk}
\affiliation{University of Aberdeen, King's College, Institute for Complex Systems and Mathematical Biology, AB24 3UE Aberdeen, United Kingdom}
\date{\today}
%
\begin{abstract}
One of the most popular and innovative methods to analyse signals is by using Ordinal Patterns (OPs). The OP encoding is based on transforming a (univariate) signal into a symbolic sequence of OPs, where each OP represents the number of permutations needed to order a small subset of the signal's magnitudes. This implies that OPs are conceptually clear, methodologically simple to implement, robust to noise, and can be applied to short signals. Moreover, they simplify the statistical analyses that can be carried out on a signal, such as entropy and complexity quantifications. However, because of the relative ordering, information about the magnitude of the signal at each timestamp is lost -- this being one of the major drawbacks in the method. Here, we propose a way to use the signal magnitudes discarded in the OP encoding as a complementary variable to its permutation entropy. To illustrate our approach, we analyse synthetic trajectories from logistic and H{\'e}non maps -- with and without added noise -- and intracranial electroencephalographic recordings from rats in different sleep-wake states. Our results show that, when complementing the permutation entropy with the variability in the signal magnitudes, the characterisation of the dynamical behaviours of the maps and the sleep-wake states is improved. This implies that our approach can be useful for feature engineering and improving AI classifiers, where typical machine learning algorithms need complementary signal features as inputs to improve classification accuracy.
\end{abstract}
\maketitle
%
\section{Introduction}\label{sec:intro}
Since the beginning of the century, the boundaries of data mining have been pushed due to the growing ability to obtain larger and more precise data sets. With increasing data availability, we need to improve how we extract, manage, and analyse data \cite{Boccaletti2016}, for example, to uncover its underlying mechanisms that generate the data or to quantify its uncertainty.

An entropy measures the average content of information, where information is understood as the degree of uncertainty in an outcome (as defined by Shannon \cite{shannon1948mathematical}). If an outcome is highly unlikely to happen, then it carries significant information because it would be surprising to record it, such as the presence of an outlier or an artifact in a signal. However, if an outcome is highly likely to happen, then it carries insignificant information because one would expect it to appear, such as a periodic signal. Hence, entropy is highest when any outcome is equally likely to happen, corresponding to a uniform probability distribution that conveys the maximum uncertainty regarding all possible outcomes \cite{shannon1948mathematical}.

One of the most successful entropy methods introduced to characterise signals is the permutation entropy \cite{bandt2002permutation}. Permutation entropy quantifies the average content of information in an Ordinal Pattern (OP) sequence, which is obtained from the signal by dividing it into a series of embedded vectors \cite{bandt2002permutation, amigo2010permutation}. Each OP represents the number of permutations needed to order the signal magnitudes within each embedded vector. The resultant symbolic sequence is used to find the OP probabilities distribution. From these probabilities, it is easier to perform statistical quantifications -- known as OP analysis -- than from the original signal, such as quantifying the uncertainty and complexity of the signal \cite{lamberti2004intensive, keller2005ordinal, zunino2022permutation}.

Due to its simplicity and robustness to noise, OP analysis (along with complexity calculations) has had remarkable success \cite{zanin2012permutation, leyva2022twenty, amigo2023ordinal}, being used to distinguish between chaotic and stochastic signals \cite{amigo2010detecting, zunino2012distinguishing, unakafov2014conditional, bandt2019small, sakellariou2019markov, zanin2021ordinal, zunino2023quantifying, kottlarz2023ordinal}, characterise electrophysiological signals \cite{quintero2018differentiating, gonzalez2019decreased, gonzalez2022low, gonzalez2023sleep, bandt2023statistics, zunino2024revisiting, boaretto2023spatial, gancio2024permutation}, laser dynamics \cite{rubido2011language, soriano2011time, aragoneses2013distinguishing, aragoneses2014unveiling, aragoneses2016unveiling, tirabassi2023permutation, boaretto2024characterizing, zunino2024identifying}, climate systems \cite{barreiro2011inferring, deza2013inferring, tupikina2014characterizing, deza2018large, dijkstra2019networks, wu2020uncovering, ruiz2021permutation, gancio2024analysis}, and financial trends \cite{zanin2008forbidden, zunino2009forbidden, zhao2013measuring, yin2014weighted, stosic2019exploring, henry2019permutation, bandt2020order, kozak2020permutation}, to name a few. However, one of the main drawbacks in OP analysis is that the magnitude of the signal at each timestamp is discarded, solely keeping the ordinal relationship between the signal magnitudes.

Here, we propose to include the standard deviation of the signal magnitudes in the OP embedded vectors as a complementary variable to the permutation entropy. 
We show that signal characterisation is improved when using these standard deviations to complement the permutation entropy analysis, where we focus mostly on the R{\'e}nyi min-entropy \cite{renyi1961measures}. Our conclusions are based on analysing numerically generated trajectories from coupled logistic \cite{may1976simple,kaneko1990clustering,l2016electronically} and H{\'e}non \cite{henon2004two} maps (with and without observational noise) and intracranial electroencephalographic recordings from rats in different sleep-wake states.

The formalism behind our approach is justified in the works of Politi \cite{politi2017quantifying, watt2019permutation}, who showed that the logarithm of the standard deviation of the signal magnitudes is needed -- along with the information dimension of the system \cite{renyi1959dimension}, which is typically difficult to obtain from experimental data -- to make the permutation entropy of a signal tend to its Kolmogorov-Sinai entropy \cite{kolmogorov1959entropy, sinai1959notion}, which is a rigorously defined observable with invariant characteristics (contrary to the permutation entropy, which can depend on the signal's length and embedding choice).
%
\section{Data, Models, and Methods}\label{sec:methods}
    \subsection{Signals: general notions}
We only consider digital signals, i.e., the time is discrete and the magnitudes are quantised. These signals can be numerically generated (synthetic) or experimentally measured, where their digital nature is due to the precision of the computer or by analog-to-digital converters, respectively. The signals we analyse come from a pair of coupled logistic maps, a H{\'e}non map -- both being synthetic bi-variate trajectories -- and intracranial electroencephalographic (EEG) recordings from rats.

We write a signal as $\{x_t\}_{t=1}^{T} = \{x_1,x_2,\ldots,x_{T}\}$, where $x_t$ is the magnitude at the discrete time index $t\in\mathbb{N}$, $x_1$ is the initial state, and $T$ is the length of the signal. A signal can be resampled using an embedding delay $\tau\in\mathbb{N}$, such that $\{x_1,x_2,\ldots,x_{T}\}\mapsto\{x_1,x_{1+\tau},\ldots,x_{1+n\tau}\}$, where $n = \lfloor (T-1)/\tau \rfloor$ is the smallest integer closest to $(T-1)/\tau$. This resampling can filter the high frequencies in a signal, but we set $\tau = 1$ for all our analyses.

    \subsection{Synthetic models: map iterates}
We generate bi-variate signals from coupled, identical, logistic maps by iterating the following equations
\begin{equation}
    \left\lbrace \begin{array}{ll}
        x_{t+1} = & \left(1-\varepsilon\right)\,f(x_t) + \varepsilon\,f(y_t),  \\
        y_{t+1} = & \left(1-\varepsilon\right)\,f(y_t) + \varepsilon\,f(x_t),
    \end{array}\right.
    \label{eq:LogMaps}
\end{equation}
where $f(z) = r\,z(1-z)$ is the logistic mapping (with $z = x_t$ or $y_t$ for $t = 1,\ldots,T$), $r\in(3,4]\subset\mathbb{R}$ is the control parameter, and $\varepsilon\in[0,1]\subset\mathbb{R}$ is the coupling strength between the maps. When $\varepsilon = 0$ in Eq.~\eqref{eq:LogMaps}, the $x$ and $y$ maps are decoupled, i.e., they are isolated. As $r$ is increased from $r = 3$ to $r = 4$, an isolated logistic map undergoes a series of period-doubling bifurcations, taking the solutions from periodic to chaotic trajectories \cite{may1976simple}. When $0 < \varepsilon \leq 1$, the maps are coupled and resultant trajectories can become more complex (including intermittent and hysterical behaviours) \cite{kaneko1990clustering,l2016electronically}.

The  H{\'e}non map is given by \cite{henon2004two}
\begin{equation}
    \left\lbrace \begin{array}{ll}
        x_{t+1} = & 1 - a\,x_t^2 + y_t,  \\
        y_{t+1} = & b\,x_t,
    \end{array}\right.
    \label{eq:Henon}
\end{equation}
where $a$ and $b$ are the control parameters, which depending on their values, can generate periodic (e.g., when $a = 1.0$ and $b = 0.3$) or chaotic (e.g., when $a = 1.4$ and $b = 0.3$) trajectories.

For both maps [Eqs.~\eqref{eq:LogMaps} and \eqref{eq:Henon}], we carry the OP analysis of the $x$ component for different control parameter values (i.e., $r$, $\varepsilon$, and $a$), fixing its length to $T = 10^6$ after removing a transient of $\delta t = 10^3$ iterations from the initial condition $x_1 = 0.65$, $y_1 = 0.44$. In this way, we discard the bi-variate nature of these maps and focus on uni-variate signals. To analyse the effects of observational noise, we add white Gaussian noise to these signals by independently drawing identically distributed random numbers from a normal distribution and changing its strength (i.e., its standard deviation).

    \subsection{Animal Model: EEG recordings}\label{sec2c}
We use EEG recordings from $11$ healthy and freely moving Wistar rats during their natural sleep-wake cycle, having food and water within the (sound attenuated and Faraday shielded) recording box. These rats have intracranially implanted electrodes, monitoring active wakefulness (AW), rapid eye movement (REM) sleep, and non-REM sleep. Details on the surgical procedure and experimental conditions can be found in Refs.~\cite{gonzalez2019decreased, gonzalez2022low, gonzalez2023sleep}. The experiments are in agreement with Uruguay's National Animal Care Law (No. 18611) and with the ``Guide to the care and use of laboratory animals'' (8th edition, National Academy Press, Washington DC, 2010). These experiments were approved by the Institutional Animal Care Committee (Comisión de Etica en el Uso de Animales), Exp.~No.~070153-000332-16.

The EEG signals we analyse are obtained by making the differences between the electrode of interest and the Cerebellum (the reference). We focus on five electrodes, those bilaterally placed above the primary motor (M1) and somatosensory (S1) cortices, plus the right olfactory bulb (OB), discarding the two electrodes from the secondary visual (V2) cortex (because they are relatively too close to the Cerebellum, increasing the relevance of the observational noise). These EEGs have a sampling frequency of $1024Hz$ and a resolution of 16 bits. To remove the degeneracies in the signal magnitudes due to the analogue-to-digital converter, we add white noise with an amplitude given by the range of the (electrode dependent) EEG times $2^{-16}$.

AW is defined by low-voltage fast waves in M1, strong theta rhythm in S1 ($4-7Hz$), and relatively high electromyographic activity. REM sleep is defined by low-voltage fast-frontal waves, a regular theta rhythm in S1, and silent electromyography (excluding occasional twitches). NREM sleep is determined by the presence of high-voltage slow-cortical waves ($1-4Hz$), sleep spindles in M1 and S1, and a reduction in electromyographic amplitudes. Additionally, a visual scoring is performed to discard artifacts and transitional states. The EEGs for these states are concatenations of artifact-free $10s$ windows that meet each state criteria. To analyse them, we fix their lengths to $T = 90\times1024$, which is the shortest length of the concatenated EEGs.

    \subsection{Method: encoding signals into ordinal patterns}
We follow Bandt-Pompe's method \cite{bandt2002permutation} to encode uni-variate signals.

First, we divide the signal into quasi-non-overlapping vectors with $D$ components, where $D>1$ is a natural number known as the embedding dimension. That is, we transform the signal $\{x_t\}_{t=1}^{T}$ to a series of vectors $\{x_1,\ldots,x_D\}$, $\{x_D,\ldots,x_{2D-1}\}$,  $\{x_{2D-1},\ldots,x_{3D-2}\}$, $\ldots$, $\{x_{(m-1)(D-1)+1},\ldots,x_{m(D-1)}\}$,
where $m = \lceil T/(D-1) \rceil$ is the largest integer closest to $T/(D-1)$. Then we transform each vector into an integer, ranging from $1$ to $D!$, according to the number of permutations needed to order their elements in increasing fashion (plus $1$). These are known as Ordinal Patterns (OPs), and the overall process encodes the signal into a symbolic sequence that preserves the local relationships between consecutive time-points but discards their magnitudes.

For example, when $D = 2$, if $x_1 < x_2$ then $\{x_1,x_2\}\mapsto 1$, because no permutations are needed to order the two-element vector. If $x_1 > x_2$, then $\{x_1,x_2\}\mapsto 2$, because we need one permutation. In total there are $D! = 2$ possible permutations between the components of the $D = 2$ embedded vector. We encode all our signals using $D = 3,\,4$, or $5$, which implies having symbolic sequences with a maximum of $D! = 6,\,24$, or $120$ different OPs, respectively. For the length $T = 90\times1024$ of the EEG signals, the average frequency of appearance of any given OP when $D = 5$ is approximately $T/D = 768$ (being higher for $D < 5$). Consequently, we have high statistical power when finding the marginal probability distribution of the OPs and the other statistical measures.

    \subsection{Statistical measures: permutation entropy, R{\'e}nyi min-entropy, and magnitude variability}
The information content of an OP is $\log\!\left[1/P(\alpha)\right]$ (as defined by Shannon \cite{shannon1948mathematical}), where $P(\alpha)$ is the probability of having OP $\alpha$ (such that $\sum_{\alpha=1}^{D!} P(\alpha) = 1$). The Shannon entropy \cite{shannon1948mathematical} of the OP sequence, $H$, is known as the permutation entropy \cite{bandt2002permutation}, and is found from
\begin{equation}
    H = \sum_{\alpha=1}^{D!} P(\alpha)\,\log_2\!\left[\frac{1}{P(\alpha)}\right] = \left\langle \log_2\!\left[\frac{1}{P(\alpha)}\right] \right\rangle,
    \label{eq:entropy}
\end{equation}
with $\langle\cdot\rangle$ the mean with respect to the probability distribution $\{P(\alpha)\}_{\alpha=1}^{D!} = \{P(1),\ldots,P(D!)\}$. The maximum value of $H$ is $H_{\max} = \log_2(D!)$, which is known as the Hartley or max-entropy and is achieved if and only if $P(\alpha) = 1/D!$ for all $\alpha$ (a uniform distribution). We use $\log_2$ in Eq.~\eqref{eq:entropy} so that the unit is the bit.

To improve the differences in the permutation entropy values of $\{P(\alpha)\}_{\alpha=1}^{D!}$ when it is close to the uniform distribution, we use the R{\'e}nyi min-entropy (in bits), $H_\infty$, which is defined by \cite{renyi1961measures}
\begin{equation}
    H_\infty = \min_{\alpha}\left\lbrace \log_2\!\left[\frac{1}{P(\alpha)}\right]\right\rbrace = -\log_2\!\left[ \max_{\alpha}\{P(\alpha)\} \right].
    \label{eq:minEnt}
\end{equation}
This means that the information content of any OP sequence is bounded between $H_{\max}$ and $H_\infty$.

\begin{figure}[htbp]
    \centering
    \includegraphics[width=1.0\linewidth]{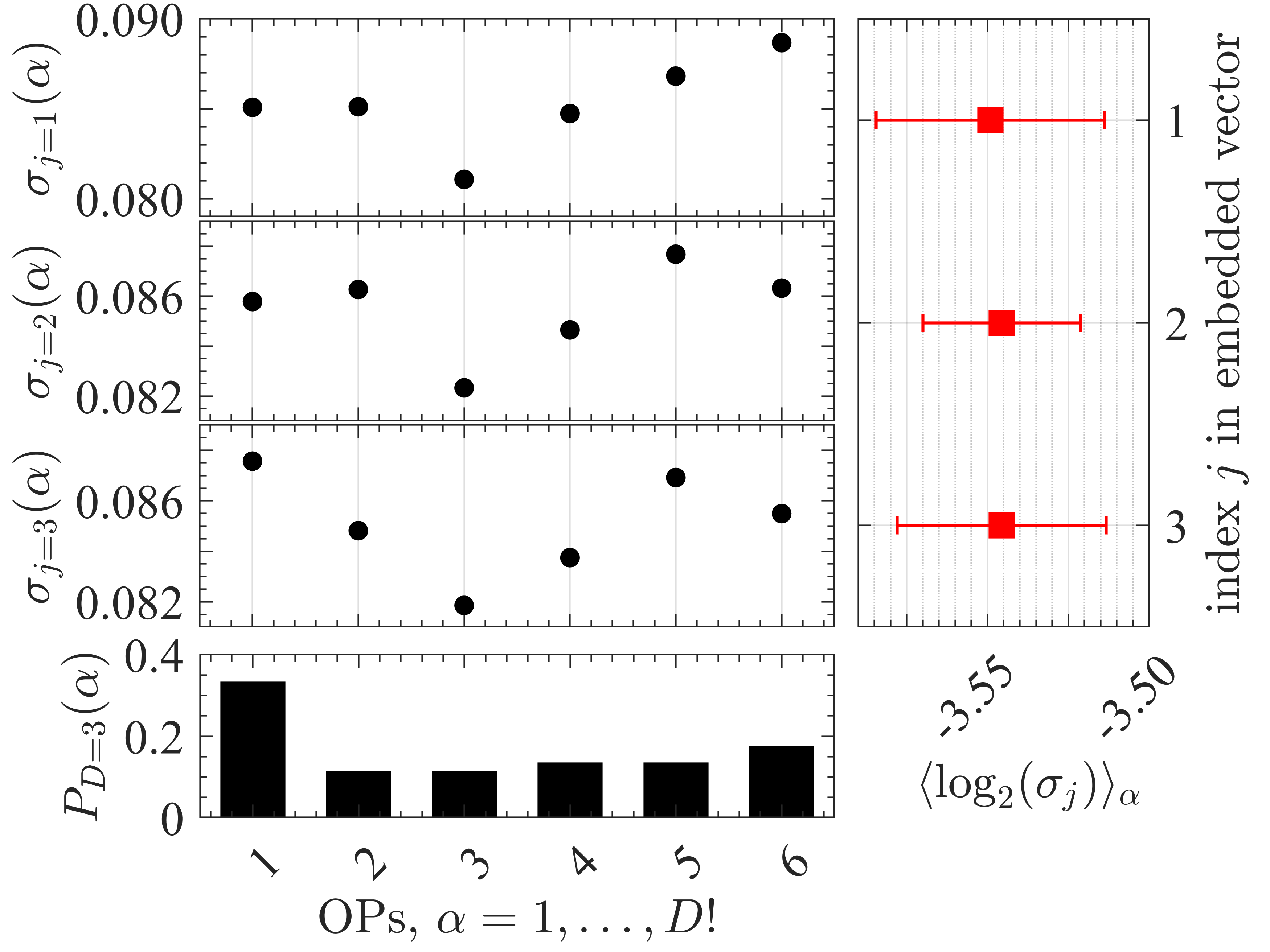}
    \caption{ {\bf Example of the variability of an EEG signal magnitudes within each ordinal pattern (OP)}. The EEG signal is from the right primary motor cortex (rM1) of a representative rat during active wakefulness and the OPs are constructed using an embedding dimension $D = 3$ and delay $\tau = 1$. The top three left panels show the values of the standard deviation of the EEG signal $\sigma_j(\alpha)$ in the components ($j = 1,2,3$) of the embedded vectors for each OP symbol ($\alpha=1,\ldots,6$). The bottom left panel shows the OP probability distribution $\{P(\alpha)\}_{\alpha=1}^{D!} = \{P(1),\ldots,P(D!=6)\}$. The right panel shows the mean (red squares) with respect to $\{P(\alpha)\}_{\alpha=1}^{6}$ for each set of $\log_2\left[\sigma_j(\alpha)\right]$ values, $\langle \log_2\left[\sigma_j\right] \rangle$, with error-bars defined by $\pm \sqrt{ \langle \log_2\left[\sigma_j\right]^2 \rangle - \langle \log_2\left[\sigma_j\right] \rangle^2 }$.}
    \label{fig_method}
\end{figure}

To quantify the variability of the signal magnitudes within the OPs, we use
\begin{equation}
    \langle \log_2\left[\sigma_j\right] \rangle = \sum_{\alpha=1}^{D!} P(\alpha) \log_2\left[\sigma_j(\alpha)\right],
    \label{eq:LogMagStd}
\end{equation}
where $\langle\cdot\rangle$ is the mean with respect to the OP probability distribution [as in Eq.~\eqref{eq:entropy}] and $\sigma_j(\alpha)$ is the standard deviation of the signal magnitudes at the $j$-th component (with $j = 1,\ldots,D$) of all the embedded vectors that correspond to the OP symbol $\alpha$ (with $\alpha = 1,\ldots,D!$) \cite{politi2017quantifying, watt2019permutation}. An example of the resultant values of Eq.~\eqref{eq:LogMagStd} for an EEG of the right primary motor cortex (rM1) of a representative rat during AW is shown in Fig.~\ref{fig_method}.

We note that the variation of the values of $\langle \log_2\left[\sigma_\alpha(j)\right] \rangle$ for different entries of $j$ is minimal, as illustrated by the example of Fig.~\ref{fig_method}. Therefore, we work with the average value (but any choice of $j$ would hold similar results and our conclusions would remain unchanged). Namely,
\begin{equation}
    \text{avg}_j\{\langle \log_2\left[\sigma_j\right] \rangle\} = \frac{1}{D}\sum_{j=1}^D \langle \log_2\left[\sigma_j\right] \rangle.
    \label{eq:avgLogMagStd}
\end{equation}

Consequently, when using Eq.~\eqref{eq:entropy} [or Eq.~\eqref{eq:minEnt}] we can quantify the average [minimum] information content in an OP sequence, but lose the information from the magnitudes that compose the embedded vectors forming the OPs. In contrast, using Eq.~\eqref{eq:avgLogMagStd} we can quantify the average magnitude variability of these embedded vectors, complementing the information provided by $H$ [or $H_\infty$].
%
\section{Results}\label{sec:results}

    \subsection{Ordinal pattern analysis of the noisless map iterates}
From the analysis of the OP sequences for the coupled logistic maps [Eq.~\eqref{eq:LogMaps}] and H{\'e}non map [Eq.~\eqref{eq:Henon}] for different parameters, we can see that when the dynamics changes slightly, the min-entropy $H_\infty$ [Eq.~\eqref{eq:minEnt}] can have small variations, but the average magnitude variability $\text{avg}_j\{\langle \log_2\left[\sigma_j\right] \rangle\}$ [Eq.~\eqref{eq:avgLogMagStd}] can change drastically.

\begin{figure}[htbp]
    \centering
    \includegraphics[width=1.0\linewidth]{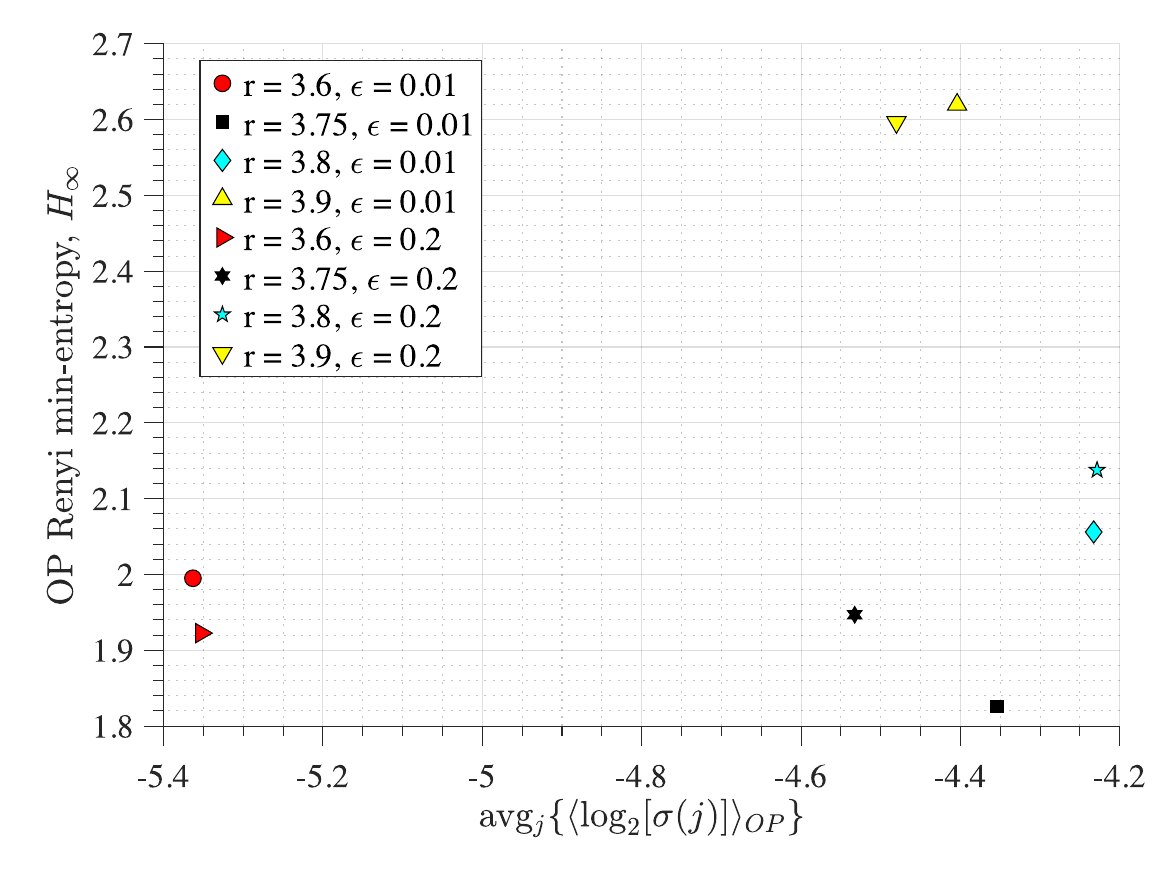}
    \caption{ {\bf R{\'e}nyi min-entropy and average magnitude variability of the ordinal pattern (OP) embedded vectors from two coupled identical logistic maps}. The map iterates for the OP encoding are obtained from Eq.~\eqref{eq:LogMaps}, where the coupling strength $\varepsilon$ is set to $0.01$ or $0.2$ and the map parameter $r$ is set to $3.6$ (red symbols), $3.75$ (black symbols), $3.8$ (cyan symbols), or $3.9$ (yellow symbols). We use $D = 4$ and $\tau = 1$ for the OP encoding of the iterates of the $x$ component (one map) -- see Sect.~\ref{sec:methods} for details. }
    \label{fig_LogMaps_H-Std}
\end{figure}

Figure~\ref{fig_LogMaps_H-Std} shows that when the coupling strength is $\varepsilon=0.01$, the red circle ($r=3.60$) and the cyan diamond ($r=3.80$) have similar $H_\infty$ values, close to $2$ and $2.05$ bits (vertical axis), respectively. In contrast, their $\text{avg}_j\{\langle \log_2\left[\sigma_j\right] \rangle\}$ values differ by an order of magnitude, being close to $-5.38$ and $-4.22$ (horizontal axis), respectively. Similarly, when $\varepsilon=0.2$, the red triangle ($r=3.60$) and the black star ($r=3.75$) have an $\text{avg}_j\{\langle \log_2\left[\sigma_j\right] \rangle\}$ differing by an order of magnitude (horizontal axis) but similar $H_\infty$ values, which are close to $1.92$ and $1.95$ bits (vertical axis), respectively.

However, the opposite effect is also observed in Fig.~\ref{fig_LogMaps_H-Std}. For example, when $\varepsilon=0.01$, the black square ($r=3.75$) and the yellow triangle pointing upward ($r=3.9$) have significantly different $H_\infty$ values (close to $1.82$ and $2.62$ bits, respectively), but their $\text{avg}_j\{\langle \log_2\left[\sigma_j\right] \rangle\}$ is similar (close to $-4.35$ and $-4.40$, respectively). The same happens for $\varepsilon=0.2$, where the black star ($r=3.75$) and the yellow triangle pointing downward ($r=3.9$) have similar $\text{avg}_j\{\langle \log_2\left[\sigma_j\right] \rangle\}$ values (close to $-4.55$ and $-4.50$, respectively) but significantly different $H_\infty$ (close to $1.94$ and $2.60$ bits, respectively).

\begin{figure}[htbp]
    \centering
    \includegraphics[width=0.95\linewidth]{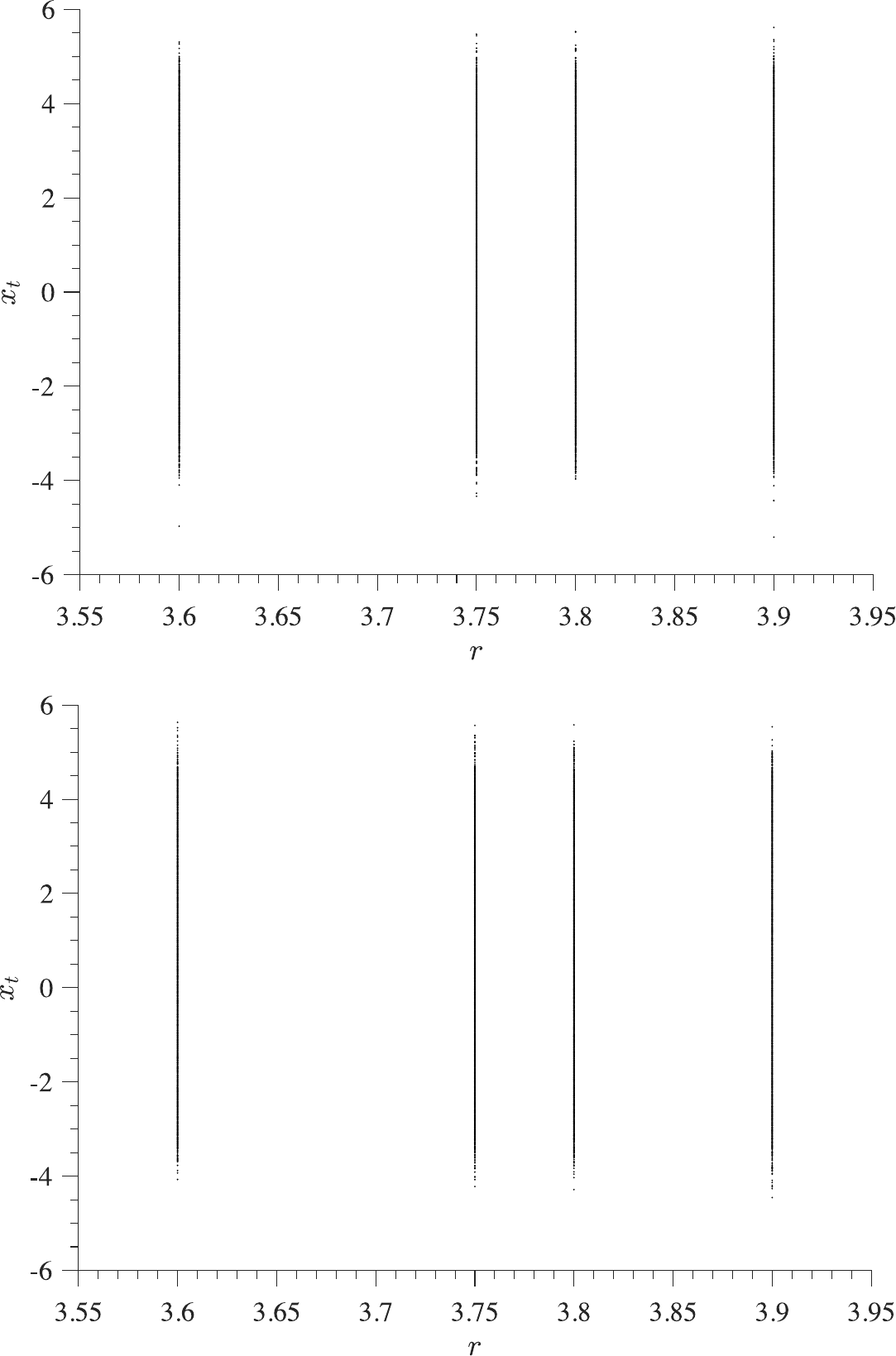}
    \caption{ {\bf Bifurcation-like diagrams for coupled identical logistic maps}. The top [bottom] panel shows the signal of one map, $x_t$ [Eq.~\eqref{eq:LogMaps}], when $\varepsilon = 0.01$ [$\varepsilon = 0.2$] as $r$ is changed according to the values used in Fig.~\ref{fig_LogMaps_H-Std}.}
    \label{fig_LogMapsBif}
\end{figure}

The dynamics of the coupled logistic maps shown in Fig.~\ref{fig_LogMaps_H-Std} follow the bifurcation-like diagram of Fig.~\ref{fig_LogMapsBif}. We can see that, whichever the dynamical changes driving the similarities (or dissimilarities) in $H_\infty$ or $\text{avg}_j\{\langle \log_2\left[\sigma_j\right] \rangle\}$ values in Fig.~\ref{fig_LogMaps_H-Std}, are unnoticeable in Fig.~\ref{fig_LogMapsBif} by doing a direct visual inspection. This implies that the dynamical changes must be happening at smaller scales than the length of the signal, which is likely why the OP encoding is able to capture the different chaoticities (i.e., different $H_\infty$ values) emerging for different control parameters and local changes in the magnitude of the $x_t$ signals (i.e., different $\text{avg}_j\{\langle \log_2\left[\sigma_j\right] \rangle\}$ values).

\begin{figure}[htbp]
    \centering
    \includegraphics[width=1.0\linewidth]{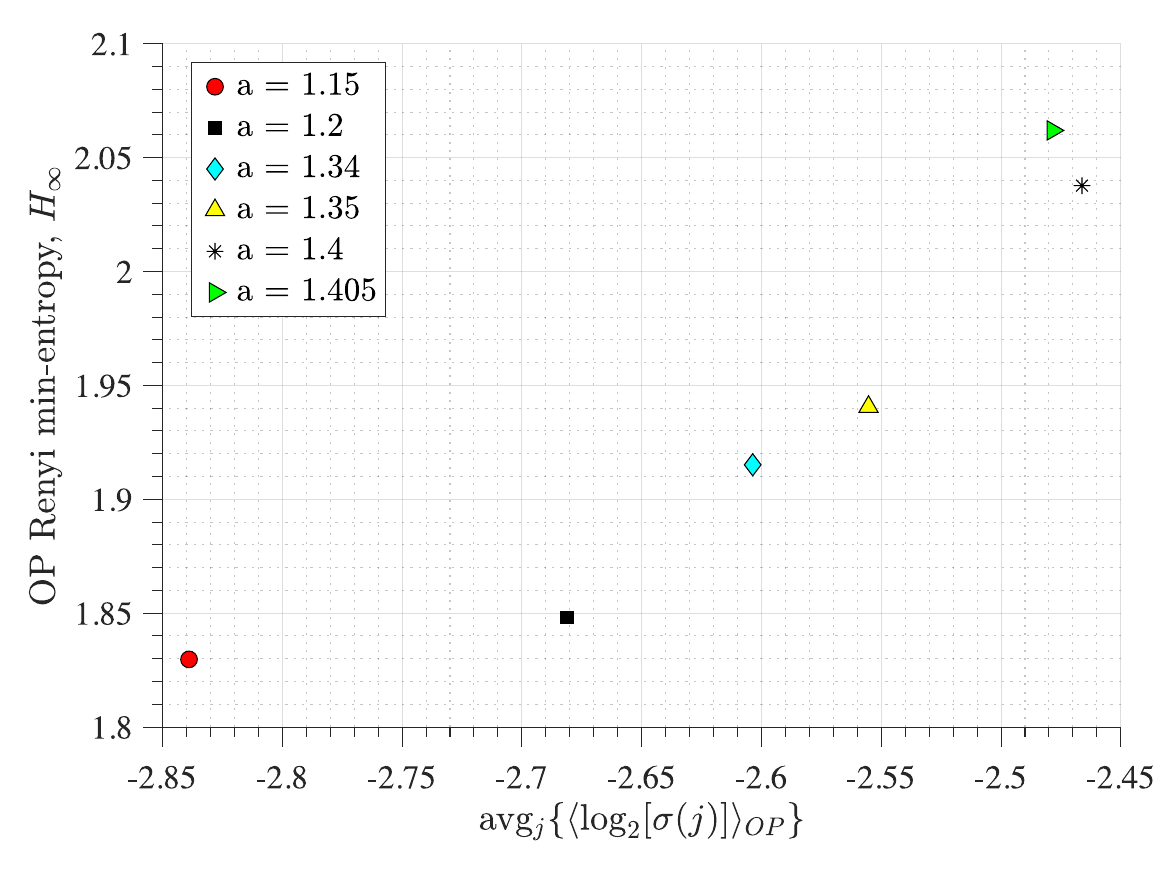}
    \caption{ {\bf R{\'e}nyi min-entropy and average magnitude variability of the ordinal pattern (OP) embedded vectors from a H{\'e}non map}. The map iterates are obtained from Eq.~\eqref{eq:Henon} with $b = 0.3$ and the other parameter set to either $a = 1.15$ (red circle), $1.20$ (black square), $1.34$ (cyan diamond), $1.35$ (yellow triangle), $1.40$ (black asterisk), or $1.405$ (green triangle). We use $D = 4$ and $\tau = 1$ for the OP encoding of the iterates of the $x$ component (as in Fig.~\ref{fig_LogMaps_H-Std}) -- see Sect.~\ref{sec:methods} for details. }
    \label{fig_Henon_H-Std}
\end{figure}

We draw similar conclusions when analysing the H{\'e}non map. For example, the red circle ($a=1.15$) and the black square ($a=1.20$) in Fig.~\ref{fig_Henon_H-Std} have similar $H_\infty$ values (close to $1.83$ and $1.85$ bits, respectively), but have different $\text{avg}_j\{\langle \log_2\left[\sigma_j\right] \rangle\}$; close to $-2.85$ for the red circle and $-2.68$ for the black square, respectively.

In these cases ($a=1.15$, $a=1.20$, $a=1.34$, $a=1.35$, $a=1.40$, and $a=1.405$), the dynamics of the map is in a chaotic regime, with apparent minimal differences (similarly to Fig.~\ref{fig_LogMapsBif}) -- this can be corroborated from the bifurcation diagram of the  H{\'e}non map (not included). However, as $a$ is increased for the values analysed, the chaoticity of the signal is also increased. This can be quantified by Lyapunov exponents or by the Kolmogorov-Sinai entropy of the system, and as Fig.~\ref{fig_Henon_H-Std} shows, it can also be quantified by using $H_\infty$ and $\text{avg}_j\{\langle \log_2\left[\sigma_j\right] \rangle\}$. 

\begin{figure}[htbp]
    \centering
    \includegraphics[width=1.0\linewidth]{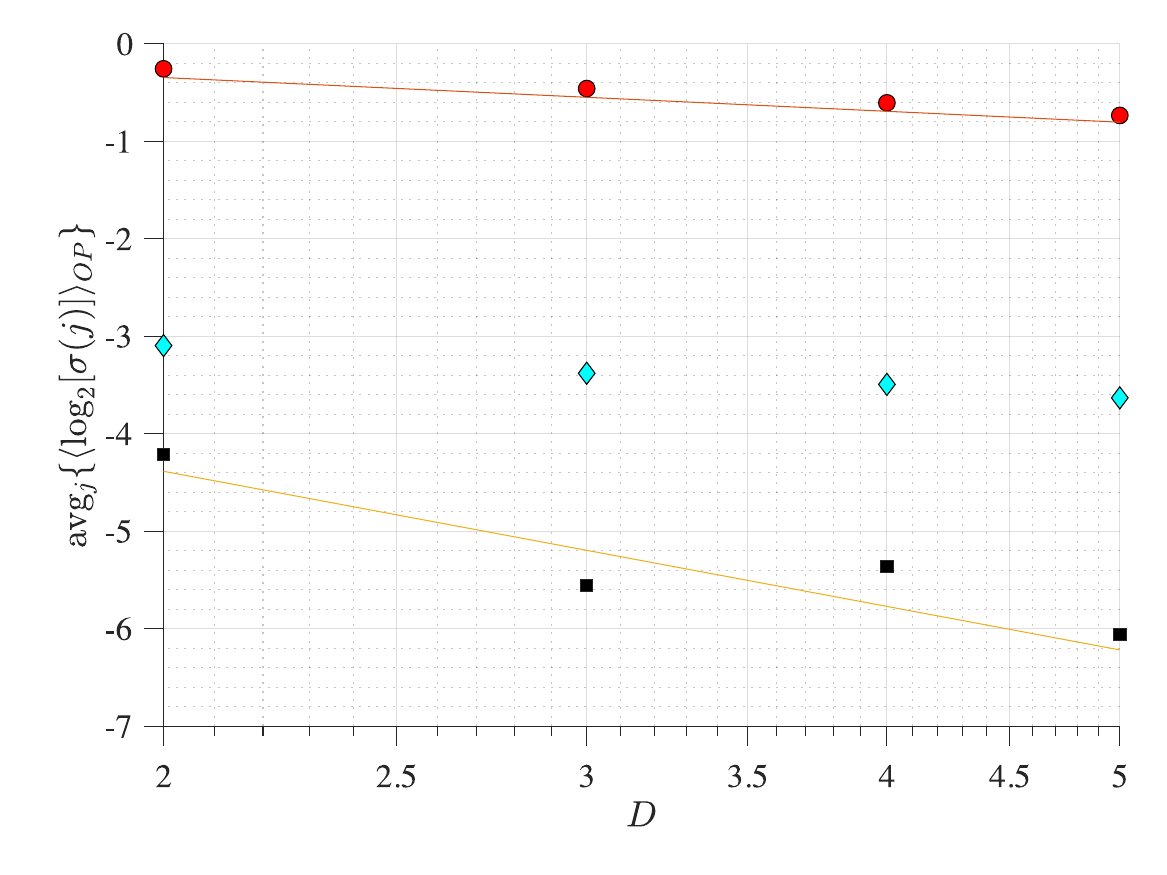}
    \caption{ {\bf Average magnitude variability of the ordinal pattern sequences from two coupled identical logistic maps as a function of the embedding dimension $D$ and noise strength}. Black squares correspond to noiseless iterates, cyan diamonds to observational noise with a standard deviation of $10^{-1}$, and red circles to observational noise with a standard deviation of $10^{0}$. The map parameters for all symbols are set such that $r=3.6$ and $\varepsilon = 0.01$ [Eq.~\eqref{eq:LogMaps}]. Two reference lines are included with slopes of $-2$ and $-1/2$.}
    \label{fig_LogMaps_Std-D}
\end{figure}

\begin{figure}[htbp]
    \centering
    \includegraphics[width=1.0\linewidth]{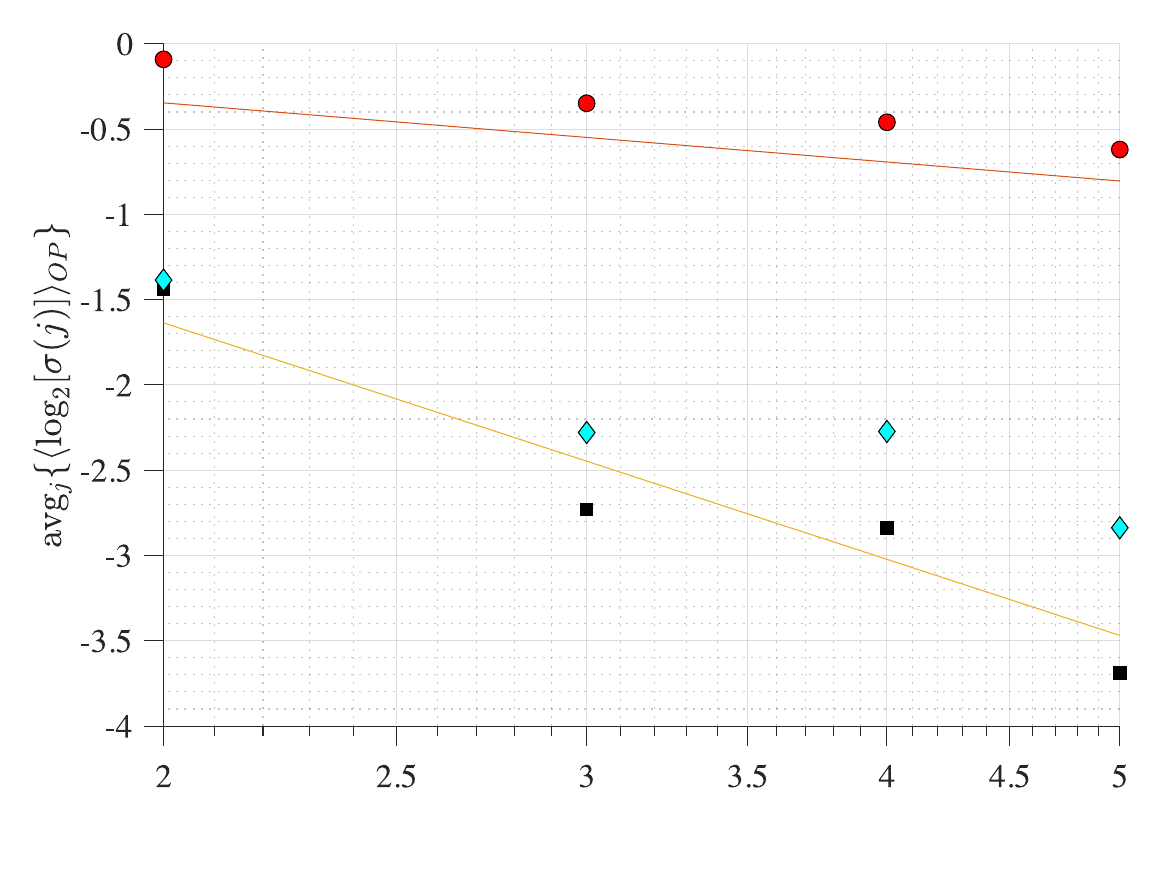}
    \caption{ {\bf Average magnitude variability of the ordinal pattern (OP) sequences from a H{\'e}non map as a function of the OP embedding dimension and noise strength}. Filled symbols have the same observational noises as in Fig.~\ref{fig_LogMaps_Std-D}. The map parameters of the H{\'e}non map are $b = 0.3$ and $a=1.15$ [Eq.~\eqref{eq:Henon}]. Two reference lines are included with slopes of $-2$ and $-1/2$.}
    \label{fig_Henon_D-Std}
\end{figure}

Overall, these results (Figs.~\ref{fig_LogMaps_H-Std} and \ref{fig_Henon_H-Std}) show that using the OP R{\'e}nyi min-entropy in conjunction with the average signal variability per OP improves the characterisation of the underlying map dynamics. Moreover, we can see that these results (Figs.~\ref{fig_LogMaps_H-Std} and \ref{fig_Henon_H-Std}) scale with the noise and the use of different embedding dimensions $D$, because $\text{avg}_j\{\langle \log_2\left[\sigma_j\right] \rangle\}$ follows a power-law behaviour as a function of $D \tau$ with an exponent that depends on the noise strength (see \cite{politi2017quantifying, watt2019permutation} for details), which can be useful to distinguish between chaotic and stochastic signals. This is corroborated in Figs.~\ref{fig_LogMaps_Std-D} and \ref{fig_Henon_D-Std} for the coupled logistic maps and H{\'e}non map, respectively.

    \subsection{Ordinal pattern analysis of the EEG recordings}
\begin{figure*}[htbp]
    \centering
    \includegraphics[width=1.0\linewidth]{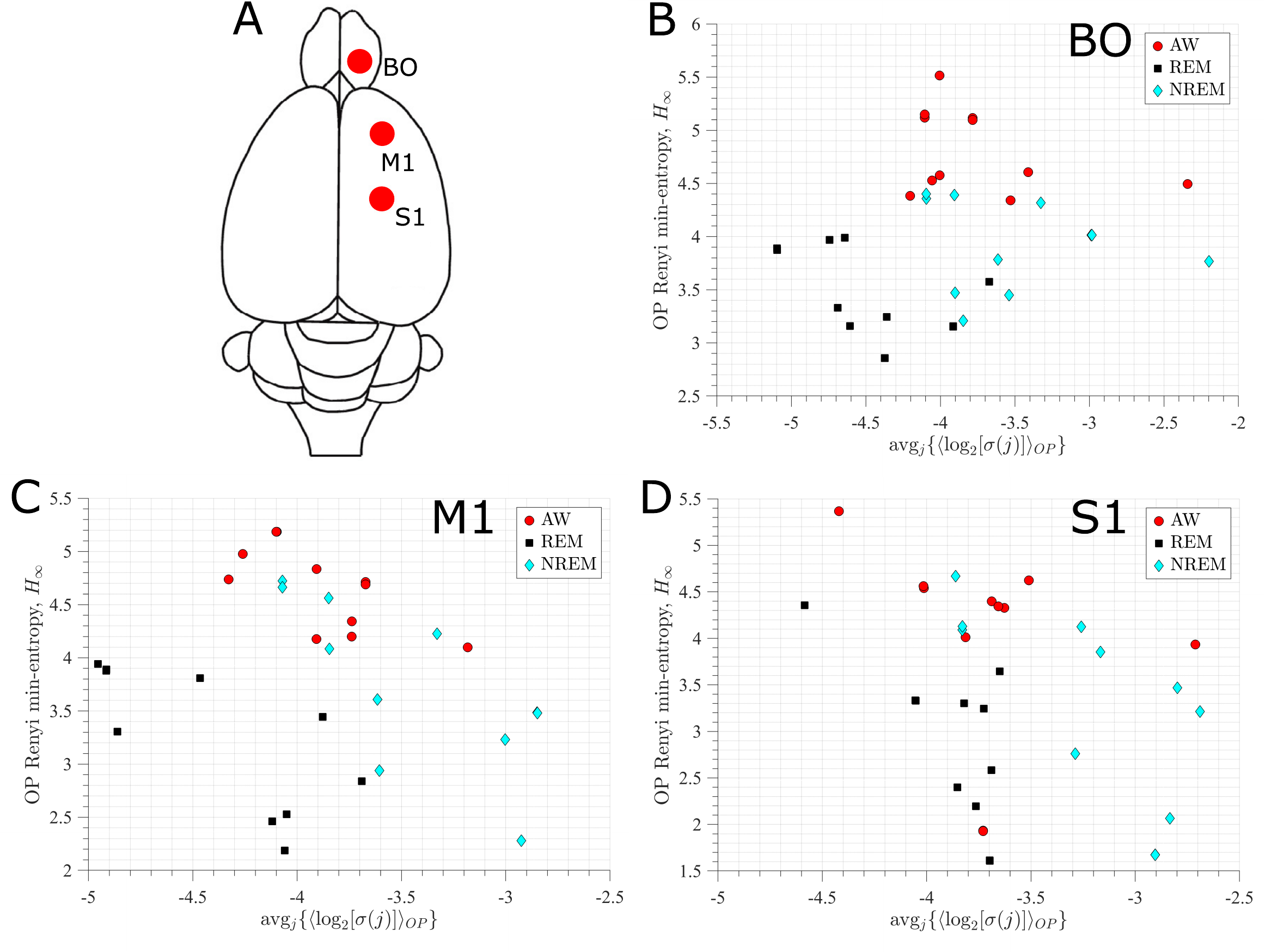}
    \caption{OPs analysis for EEG recordings from 11 rats in three sleep-wake states: active wakefulness (AW), rapid eye movement (REM) and non-REM (NREM) sleep. (A) Location of the electrodes for the EEG recordings corresponding to the right olfactory bulb (BO), the primary motor (M1) and somatosensory (S1) cortices. (B), (C), (D) Rényi entropy vs. average magnitude variability, respectively, for BO, M1 and S1 electrodes. The embedding dimension for the OPs is $D=5$\,.}
    \label{fig_EEGs}
\end{figure*}

The sleep-wake states of active wakefulness (AW), rapid-eye movement (REM) sleep, and non-REM (NREM) sleep have different electrophysiological characteristics (see Sect.~\ref{sec2c}). However, when focusing solely on frequency or permutation entropy analyses, some of these differences are lost \cite{gonzalez2019decreased, gonzalez2022low, gonzalez2023sleep}. Here, we show that using the average signal variability per OP ($\text{avg}_j\{\langle \log_2\left[\sigma_j\right] \rangle\}$) improves the differentiation between the states, even when accounting for natural inter-animal variability.

The results of the EEGs for the three sleep-wake states coming from the right OB, M1, and S1 of $11$ rats are shown in Fig.~\ref{fig_EEGs} -- we find similar results for the left hemisphere's M1 and S1 cortices. The panels {\bf B} (OB), {\bf C} (M1), and {\bf D} (S1) show that AW has the highest value of $H_\infty$ for all animals -- with one exception, in the S1 cortex (panel {\bf D}, red filled circles) -- but mid-range values of $\text{avg}_j\{\langle \log_2\left[\sigma_j\right] \rangle\}$. This implies that the permutation entropy can distinguish considerably well between wakefulness and sleep states. However, the values of $H_\infty$ for REM and NREM are similar for all electrodes.

In contrast, the values of $\text{avg}_j\{\langle \log_2\left[\sigma_j\right] \rangle\}$ for REM and NREM sleep differ by approximately an order of magnitude in most animals. This is in accordance with the type of waves present in these sleep states: REM has low-voltage fast-frontal waves but NREM has high-voltage slow-cortical waves (see Sect.~\ref{sec2c}). Consequently, the AW, REM, and NREM states are fairly differentiated when using $H_\infty$ and $\text{avg}_j\{\langle \log_2\left[\sigma_j\right] \rangle\}$ simultaneously. We note that this differentiation is improved as the cortical location considered is further away from the reference electrode, which in our case is the right OB and Cerebellum, respectively. We also note that these results and conclusions remain unchanged when using $H$ [Eq.~\eqref{eq:entropy}] instead of $H_\infty$ [Eq.~\eqref{eq:minEnt}].

\section{Discussion}\label{sec:discuss}
For many real-world systems, the intrinsic dynamics is so complex that inferring a mathematical model for the underlying dynamical process generating signal measurements becomes a very challenging task -- if feasible at all. An alternative approach is to analyse the evolution of the information contained in the signals one can measure. While this approach fails to provide a detailed model for the microscopic dynamics, if successful, then it allows to characterise different dynamical regimes and changes in the parameters of the system.

With this in mind, here we propose to include the magnitude variability of the signal in the ordinal pattern (OP) encoding. We do this to complement the permutation entropy analysis and improve the characterisation of the dynamical behaviours observed from time-series.

We first test our approach on synthetically generated signals from chaotic bidimensional mappings under different parameter values. Specifically, we use coupled logistic maps [Eq.~\eqref{eq:LogMaps}] and the H\'enon map [Eq.~\eqref{eq:Henon}]. We analyse the signals coming from one of the two coordinates available for these maps because the OP encoding can only be applied to univariate signals.

The main reason to choose bidimensional mappings is that there are proofs showing that the permutation entropy of one-dimensional mappings directly relates to the Kolmogorov-Sinai entropy \cite{bandt2002entropy, amigo2005permutation}, which is not the case for higher-dimensional mappings. Because of this difference, it is expected that the characterisation of a dynamical regime by the permutation entropy for a higher dimensional system is incomplete. We show that this problem is mitigated by including the standard deviation of the signal (in the embedded vectors forming the OPs) to the permutation entropy quantification (Figs.~\ref{fig_LogMaps_H-Std} and \ref{fig_Henon_H-Std}). We also find that our approach improves the characterisation of the dynamical regimes even under significant levels of observational noise and different choices of embedding dimension (Figs.~\ref{fig_LogMaps_Std-D} and \ref{fig_Henon_D-Std}).

We then analyse real-world EEG signals registered intracranially from $11$ rats under free conditions through the sleep-wake cycle. We consider these signals because they come from a system -- the brain -- where the underlying microscopic dynamics is unknown (i.e., we lack a differential equation that models the system), there are inherent noise sources affecting the quality of the signal measurements, and it has been hypothesised to have some level of chaoticity. Moreover, in practice, the polysomnographic classification of sleep-wake states requires highly trained professionals to recognise characteristic electrophysiological patterns that vary according to the sleep stage, plus depend on anatomy and individual variability. However, the electrophysiological variability introduced by experimental manipulations (in research settings) or disease (in clinical settings), requires that whichever automatic sleep scoring classifier used be interpretable and not a black box.

Our results show that, for most cortical locations, independently of the embedding dimension used, and the natural inter-animal variability, the states of active wakefulness, rapid-eye movement (REM) sleep, and non-REM sleep can be distinguished in the plane formed by the permutation entropy and signal variability  (Fig.~\ref{fig_EEGs}).

Our analyses are limited in terms of the choice of embedding delay and overlap between consecutive embedded vectors; namely, for all our results $\tau = 1$ and the embedded vectors are quasi-non-overlapping, only sharing a single data point from the signal. The reason to restrict $\tau$ to $1$ is that we can consider the entire signal, whereas $\tau > 1$ implies considering sub-sampled signals. On the other hand, increasing the overlapping of the embedded vectors creates artificial correlations between consecutive OPs (which are irrelevant for most permutation entropy calculations, but can affect conditional or transfer entropy calculations). For example, with a larger overlap than the one we choose here, two consecutive embedded vectors with $D = 3$ would share two points, such that $\{x_{t},x_{t+1},x_{t+2}\}$ and $\{x_{t+1},x_{t+2},x_{t+3}\}$. This would imply that, if $x_{t+1} > x_{t+2}$, then the OP for the second embedded vector would be conditioned to having this increasing relation that is present in the previous embedded vector, and all other OP possibilities would be forbidden (i.e., the OP would be artificially forced to take on particular OP symbols). Consequently, our choice of encoding parameters allows to keep all the signal and have a maximum number of embedded vectors with null redundancy between consecutive ones.

Finally, we note that a natural extension to our approach could consider a three-dimensional space, composed of the permutation entropy, signal variability within the embedded vectors, and a complexity measure, such as the Jensen-Shannon complexity, which could improve the characterisation of other dynamical regimes (such as non-chaotic ones). Moreover, in practical applications where the signals are short, instead of considering the standard deviation of the signal for the embedded vectors, one could consider the inter-quartile range, which is a statistically robust descriptor that is unaffected by outliers.
%
\section{Author Contributions}
MT: Investigation, Formal Analysis, Software, Visualization, Writing (original draft).
JG: Data Acquisition, Investigation.
NR: Conceptualization, Methodology, Software, Writing (original draft).
All authors: Writing - review \& editing.
%
\section{Acknowledgments}
We thank the Departamento de Fisiolog{\'i}a de Facultad de Medicina, Universidad de la Rep{\'u}blica, Montevideo, Uruguay, for providing the data set, and in particular, we thank Drs Mat{\'i}as Cavelli, Santiago Castro-Zaballa, Alejandra Mondino, and Prof Pablo Torterolo.
MT acknowledges the support of the UKRI grant MR/X034240/1.


\begin{thebibliography}{62}%
\makeatletter
\providecommand \@ifxundefined [1]{%
 \@ifx{#1\undefined}
}%
\providecommand \@ifnum [1]{%
 \ifnum #1\expandafter \@firstoftwo
 \else \expandafter \@secondoftwo
 \fi
}%
\providecommand \@ifx [1]{%
 \ifx #1\expandafter \@firstoftwo
 \else \expandafter \@secondoftwo
 \fi
}%
\providecommand \natexlab [1]{#1}%
\providecommand \enquote  [1]{``#1''}%
\providecommand \bibnamefont  [1]{#1}%
\providecommand \bibfnamefont [1]{#1}%
\providecommand \citenamefont [1]{#1}%
\providecommand \href@noop [0]{\@secondoftwo}%
\providecommand \href [0]{\begingroup \@sanitize@url \@href}%
\providecommand \@href[1]{\@@startlink{#1}\@@href}%
\providecommand \@@href[1]{\endgroup#1\@@endlink}%
\providecommand \@sanitize@url [0]{\catcode `\\12\catcode `\$12\catcode `\&12\catcode `\#12\catcode `\^12\catcode `\_12\catcode `\%12\relax}%
\providecommand \@@startlink[1]{}%
\providecommand \@@endlink[0]{}%
\providecommand \url  [0]{\begingroup\@sanitize@url \@url }%
\providecommand \@url [1]{\endgroup\@href {#1}{\urlprefix }}%
\providecommand \urlprefix  [0]{URL }%
\providecommand \Eprint [0]{\href }%
\providecommand \doibase [0]{https://doi.org/}%
\providecommand \selectlanguage [0]{\@gobble}%
\providecommand \bibinfo  [0]{\@secondoftwo}%
\providecommand \bibfield  [0]{\@secondoftwo}%
\providecommand \translation [1]{[#1]}%
\providecommand \BibitemOpen [0]{}%
\providecommand \bibitemStop [0]{}%
\providecommand \bibitemNoStop [0]{.\EOS\space}%
\providecommand \EOS [0]{\spacefactor3000\relax}%
\providecommand \BibitemShut  [1]{\csname bibitem#1\endcsname}%
\let\auto@bib@innerbib\@empty
\bibitem [{\citenamefont {Zanin}\ \emph {et~al.}(2016)\citenamefont {Zanin}, \citenamefont {Papo}, \citenamefont {Sousa}, \citenamefont {Menasalvas}, \citenamefont {Nicchi}, \citenamefont {Kubik},\ and\ \citenamefont {Boccaletti}}]{Boccaletti2016}%
  \BibitemOpen
  \bibfield  {author} {\bibinfo {author} {\bibfnamefont {M.}~\bibnamefont {Zanin}}, \bibinfo {author} {\bibfnamefont {D.}~\bibnamefont {Papo}}, \bibinfo {author} {\bibfnamefont {P.~A.}\ \bibnamefont {Sousa}}, \bibinfo {author} {\bibfnamefont {E.}~\bibnamefont {Menasalvas}}, \bibinfo {author} {\bibfnamefont {A.}~\bibnamefont {Nicchi}}, \bibinfo {author} {\bibfnamefont {E.}~\bibnamefont {Kubik}},\ and\ \bibinfo {author} {\bibfnamefont {S.}~\bibnamefont {Boccaletti}},\ }\href@noop {} {\bibfield  {journal} {\bibinfo  {journal} {Phys. Rep.}\ }\textbf {\bibinfo {volume} {635}},\ \bibinfo {pages} {1} (\bibinfo {year} {2016})}\BibitemShut {NoStop}%
\bibitem [{\citenamefont {Shannon}(1948)}]{shannon1948mathematical}%
  \BibitemOpen
  \bibfield  {author} {\bibinfo {author} {\bibfnamefont {C.~E.}\ \bibnamefont {Shannon}},\ }\href@noop {} {\bibfield  {journal} {\bibinfo  {journal} {The Bell system technical journal}\ }\textbf {\bibinfo {volume} {27}},\ \bibinfo {pages} {379} (\bibinfo {year} {1948})}\BibitemShut {NoStop}%
\bibitem [{\citenamefont {Bandt}\ and\ \citenamefont {Pompe}(2002)}]{bandt2002permutation}%
  \BibitemOpen
  \bibfield  {author} {\bibinfo {author} {\bibfnamefont {C.}~\bibnamefont {Bandt}}\ and\ \bibinfo {author} {\bibfnamefont {B.}~\bibnamefont {Pompe}},\ }\href@noop {} {\bibfield  {journal} {\bibinfo  {journal} {Physical review letters}\ }\textbf {\bibinfo {volume} {88}},\ \bibinfo {pages} {174102} (\bibinfo {year} {2002})}\BibitemShut {NoStop}%
\bibitem [{\citenamefont {Amig{\'o}}(2010)}]{amigo2010permutation}%
  \BibitemOpen
  \bibfield  {author} {\bibinfo {author} {\bibfnamefont {J.}~\bibnamefont {Amig{\'o}}},\ }\href@noop {} {\emph {\bibinfo {title} {Permutation complexity in dynamical systems: ordinal patterns, permutation entropy and all that}}}\ (\bibinfo  {publisher} {Springer Science \& Business Media},\ \bibinfo {year} {2010})\BibitemShut {NoStop}%
\bibitem [{\citenamefont {Lamberti}\ \emph {et~al.}(2004)\citenamefont {Lamberti}, \citenamefont {Martin}, \citenamefont {Plastino},\ and\ \citenamefont {Rosso}}]{lamberti2004intensive}%
  \BibitemOpen
  \bibfield  {author} {\bibinfo {author} {\bibfnamefont {P.}~\bibnamefont {Lamberti}}, \bibinfo {author} {\bibfnamefont {M.}~\bibnamefont {Martin}}, \bibinfo {author} {\bibfnamefont {A.}~\bibnamefont {Plastino}},\ and\ \bibinfo {author} {\bibfnamefont {O.}~\bibnamefont {Rosso}},\ }\href@noop {} {\bibfield  {journal} {\bibinfo  {journal} {Physica A: Statistical Mechanics and its Applications}\ }\textbf {\bibinfo {volume} {334}},\ \bibinfo {pages} {119} (\bibinfo {year} {2004})}\BibitemShut {NoStop}%
\bibitem [{\citenamefont {Keller}\ and\ \citenamefont {Sinn}(2005)}]{keller2005ordinal}%
  \BibitemOpen
  \bibfield  {author} {\bibinfo {author} {\bibfnamefont {K.}~\bibnamefont {Keller}}\ and\ \bibinfo {author} {\bibfnamefont {M.}~\bibnamefont {Sinn}},\ }\href@noop {} {\bibfield  {journal} {\bibinfo  {journal} {Physica A: Statistical Mechanics and its Applications}\ }\textbf {\bibinfo {volume} {356}},\ \bibinfo {pages} {114} (\bibinfo {year} {2005})}\BibitemShut {NoStop}%
\bibitem [{\citenamefont {Zunino}\ \emph {et~al.}(2022)\citenamefont {Zunino}, \citenamefont {Olivares}, \citenamefont {Ribeiro},\ and\ \citenamefont {Rosso}}]{zunino2022permutation}%
  \BibitemOpen
  \bibfield  {author} {\bibinfo {author} {\bibfnamefont {L.}~\bibnamefont {Zunino}}, \bibinfo {author} {\bibfnamefont {F.}~\bibnamefont {Olivares}}, \bibinfo {author} {\bibfnamefont {H.~V.}\ \bibnamefont {Ribeiro}},\ and\ \bibinfo {author} {\bibfnamefont {O.~A.}\ \bibnamefont {Rosso}},\ }\href@noop {} {\bibfield  {journal} {\bibinfo  {journal} {Physical Review E}\ }\textbf {\bibinfo {volume} {105}},\ \bibinfo {pages} {045310} (\bibinfo {year} {2022})}\BibitemShut {NoStop}%
\bibitem [{\citenamefont {Zanin}\ \emph {et~al.}(2012)\citenamefont {Zanin}, \citenamefont {Zunino}, \citenamefont {Rosso},\ and\ \citenamefont {Papo}}]{zanin2012permutation}%
  \BibitemOpen
  \bibfield  {author} {\bibinfo {author} {\bibfnamefont {M.}~\bibnamefont {Zanin}}, \bibinfo {author} {\bibfnamefont {L.}~\bibnamefont {Zunino}}, \bibinfo {author} {\bibfnamefont {O.~A.}\ \bibnamefont {Rosso}},\ and\ \bibinfo {author} {\bibfnamefont {D.}~\bibnamefont {Papo}},\ }\href@noop {} {\bibfield  {journal} {\bibinfo  {journal} {Entropy}\ }\textbf {\bibinfo {volume} {14}},\ \bibinfo {pages} {1553} (\bibinfo {year} {2012})}\BibitemShut {NoStop}%
\bibitem [{\citenamefont {Leyva}\ \emph {et~al.}(2022)\citenamefont {Leyva}, \citenamefont {Mart{\'\i}nez}, \citenamefont {Masoller}, \citenamefont {Rosso},\ and\ \citenamefont {Zanin}}]{leyva2022twenty}%
  \BibitemOpen
  \bibfield  {author} {\bibinfo {author} {\bibfnamefont {I.}~\bibnamefont {Leyva}}, \bibinfo {author} {\bibfnamefont {J.~H.}\ \bibnamefont {Mart{\'\i}nez}}, \bibinfo {author} {\bibfnamefont {C.}~\bibnamefont {Masoller}}, \bibinfo {author} {\bibfnamefont {O.~A.}\ \bibnamefont {Rosso}},\ and\ \bibinfo {author} {\bibfnamefont {M.}~\bibnamefont {Zanin}},\ }\href@noop {} {\bibfield  {journal} {\bibinfo  {journal} {Europhysics Letters}\ }\textbf {\bibinfo {volume} {138}},\ \bibinfo {pages} {31001} (\bibinfo {year} {2022})}\BibitemShut {NoStop}%
\bibitem [{\citenamefont {Amig{\'o}}\ and\ \citenamefont {Rosso}(2023)}]{amigo2023ordinal}%
  \BibitemOpen
  \bibfield  {author} {\bibinfo {author} {\bibfnamefont {J.~M.}\ \bibnamefont {Amig{\'o}}}\ and\ \bibinfo {author} {\bibfnamefont {O.~A.}\ \bibnamefont {Rosso}},\ }\href@noop {} {\bibfield  {journal} {\bibinfo  {journal} {Chaos: An Interdisciplinary Journal of Nonlinear Science}\ }\textbf {\bibinfo {volume} {33}},\ \bibinfo {pages} {080401} (\bibinfo {year} {2023})}\BibitemShut {NoStop}%
\bibitem [{\citenamefont {Amig{\'o}}\ \emph {et~al.}(2010)\citenamefont {Amig{\'o}}, \citenamefont {Zambrano},\ and\ \citenamefont {Sanju{\'a}n}}]{amigo2010detecting}%
  \BibitemOpen
  \bibfield  {author} {\bibinfo {author} {\bibfnamefont {J.~M.}\ \bibnamefont {Amig{\'o}}}, \bibinfo {author} {\bibfnamefont {S.}~\bibnamefont {Zambrano}},\ and\ \bibinfo {author} {\bibfnamefont {M.~A.}\ \bibnamefont {Sanju{\'a}n}},\ }\href@noop {} {\bibfield  {journal} {\bibinfo  {journal} {International Journal of Bifurcation and Chaos}\ }\textbf {\bibinfo {volume} {20}},\ \bibinfo {pages} {2915} (\bibinfo {year} {2010})}\BibitemShut {NoStop}%
\bibitem [{\citenamefont {Zunino}\ \emph {et~al.}(2012)\citenamefont {Zunino}, \citenamefont {Soriano},\ and\ \citenamefont {Rosso}}]{zunino2012distinguishing}%
  \BibitemOpen
  \bibfield  {author} {\bibinfo {author} {\bibfnamefont {L.}~\bibnamefont {Zunino}}, \bibinfo {author} {\bibfnamefont {M.~C.}\ \bibnamefont {Soriano}},\ and\ \bibinfo {author} {\bibfnamefont {O.~A.}\ \bibnamefont {Rosso}},\ }\href@noop {} {\bibfield  {journal} {\bibinfo  {journal} {Physical Review E—Statistical, Nonlinear, and Soft Matter Physics}\ }\textbf {\bibinfo {volume} {86}},\ \bibinfo {pages} {046210} (\bibinfo {year} {2012})}\BibitemShut {NoStop}%
\bibitem [{\citenamefont {Unakafov}\ and\ \citenamefont {Keller}(2014)}]{unakafov2014conditional}%
  \BibitemOpen
  \bibfield  {author} {\bibinfo {author} {\bibfnamefont {A.~M.}\ \bibnamefont {Unakafov}}\ and\ \bibinfo {author} {\bibfnamefont {K.}~\bibnamefont {Keller}},\ }\href@noop {} {\bibfield  {journal} {\bibinfo  {journal} {Physica D: Nonlinear Phenomena}\ }\textbf {\bibinfo {volume} {269}},\ \bibinfo {pages} {94} (\bibinfo {year} {2014})}\BibitemShut {NoStop}%
\bibitem [{\citenamefont {Bandt}(2019)}]{bandt2019small}%
  \BibitemOpen
  \bibfield  {author} {\bibinfo {author} {\bibfnamefont {C.}~\bibnamefont {Bandt}},\ }\href@noop {} {\bibfield  {journal} {\bibinfo  {journal} {Entropy}\ }\textbf {\bibinfo {volume} {21}},\ \bibinfo {pages} {613} (\bibinfo {year} {2019})}\BibitemShut {NoStop}%
\bibitem [{\citenamefont {Sakellariou}\ \emph {et~al.}(2019)\citenamefont {Sakellariou}, \citenamefont {Stemler},\ and\ \citenamefont {Small}}]{sakellariou2019markov}%
  \BibitemOpen
  \bibfield  {author} {\bibinfo {author} {\bibfnamefont {K.}~\bibnamefont {Sakellariou}}, \bibinfo {author} {\bibfnamefont {T.}~\bibnamefont {Stemler}},\ and\ \bibinfo {author} {\bibfnamefont {M.}~\bibnamefont {Small}},\ }\href@noop {} {\bibfield  {journal} {\bibinfo  {journal} {Physical Review E}\ }\textbf {\bibinfo {volume} {100}},\ \bibinfo {pages} {062307} (\bibinfo {year} {2019})}\BibitemShut {NoStop}%
\bibitem [{\citenamefont {Zanin}\ and\ \citenamefont {Olivares}(2021)}]{zanin2021ordinal}%
  \BibitemOpen
  \bibfield  {author} {\bibinfo {author} {\bibfnamefont {M.}~\bibnamefont {Zanin}}\ and\ \bibinfo {author} {\bibfnamefont {F.}~\bibnamefont {Olivares}},\ }\href@noop {} {\bibfield  {journal} {\bibinfo  {journal} {Communications Physics}\ }\textbf {\bibinfo {volume} {4}},\ \bibinfo {pages} {190} (\bibinfo {year} {2021})}\BibitemShut {NoStop}%
\bibitem [{\citenamefont {Zunino}\ and\ \citenamefont {Soriano}(2023)}]{zunino2023quantifying}%
  \BibitemOpen
  \bibfield  {author} {\bibinfo {author} {\bibfnamefont {L.}~\bibnamefont {Zunino}}\ and\ \bibinfo {author} {\bibfnamefont {M.~C.}\ \bibnamefont {Soriano}},\ }\href@noop {} {\bibfield  {journal} {\bibinfo  {journal} {Physical Review E}\ }\textbf {\bibinfo {volume} {108}},\ \bibinfo {pages} {065302} (\bibinfo {year} {2023})}\BibitemShut {NoStop}%
\bibitem [{\citenamefont {Kottlarz}\ and\ \citenamefont {Parlitz}(2023)}]{kottlarz2023ordinal}%
  \BibitemOpen
  \bibfield  {author} {\bibinfo {author} {\bibfnamefont {I.}~\bibnamefont {Kottlarz}}\ and\ \bibinfo {author} {\bibfnamefont {U.}~\bibnamefont {Parlitz}},\ }\href@noop {} {\bibfield  {journal} {\bibinfo  {journal} {Chaos: An Interdisciplinary Journal of Nonlinear Science}\ }\textbf {\bibinfo {volume} {33}},\ \bibinfo {pages} {053105} (\bibinfo {year} {2023})}\BibitemShut {NoStop}%
\bibitem [{\citenamefont {Quintero-Quiroz}\ \emph {et~al.}(2018)\citenamefont {Quintero-Quiroz}, \citenamefont {Montesano}, \citenamefont {Pons}, \citenamefont {Torrent}, \citenamefont {Garc{\'\i}a-Ojalvo},\ and\ \citenamefont {Masoller}}]{quintero2018differentiating}%
  \BibitemOpen
  \bibfield  {author} {\bibinfo {author} {\bibfnamefont {C.}~\bibnamefont {Quintero-Quiroz}}, \bibinfo {author} {\bibfnamefont {L.}~\bibnamefont {Montesano}}, \bibinfo {author} {\bibfnamefont {A.~J.}\ \bibnamefont {Pons}}, \bibinfo {author} {\bibfnamefont {M.~C.}\ \bibnamefont {Torrent}}, \bibinfo {author} {\bibfnamefont {J.}~\bibnamefont {Garc{\'\i}a-Ojalvo}},\ and\ \bibinfo {author} {\bibfnamefont {C.}~\bibnamefont {Masoller}},\ }\href@noop {} {\bibfield  {journal} {\bibinfo  {journal} {Chaos: An Interdisciplinary Journal of Nonlinear Science}\ }\textbf {\bibinfo {volume} {28}},\ \bibinfo {pages} {106307} (\bibinfo {year} {2018})}\BibitemShut {NoStop}%
\bibitem [{\citenamefont {Gonz{\'a}lez}\ \emph {et~al.}(2019)\citenamefont {Gonz{\'a}lez}, \citenamefont {Cavelli}, \citenamefont {Mondino}, \citenamefont {Pascovich}, \citenamefont {Castro-Zaballa}, \citenamefont {Torterolo},\ and\ \citenamefont {Rubido}}]{gonzalez2019decreased}%
  \BibitemOpen
  \bibfield  {author} {\bibinfo {author} {\bibfnamefont {J.}~\bibnamefont {Gonz{\'a}lez}}, \bibinfo {author} {\bibfnamefont {M.}~\bibnamefont {Cavelli}}, \bibinfo {author} {\bibfnamefont {A.}~\bibnamefont {Mondino}}, \bibinfo {author} {\bibfnamefont {C.}~\bibnamefont {Pascovich}}, \bibinfo {author} {\bibfnamefont {S.}~\bibnamefont {Castro-Zaballa}}, \bibinfo {author} {\bibfnamefont {P.}~\bibnamefont {Torterolo}},\ and\ \bibinfo {author} {\bibfnamefont {N.}~\bibnamefont {Rubido}},\ }\href@noop {} {\bibfield  {journal} {\bibinfo  {journal} {Scientific reports}\ }\textbf {\bibinfo {volume} {9}},\ \bibinfo {pages} {18457} (\bibinfo {year} {2019})}\BibitemShut {NoStop}%
\bibitem [{\citenamefont {Gonz{\'a}lez}\ \emph {et~al.}(2022)\citenamefont {Gonz{\'a}lez}, \citenamefont {Mateos}, \citenamefont {Cavelli}, \citenamefont {Mondino}, \citenamefont {Pascovich}, \citenamefont {Torterolo},\ and\ \citenamefont {Rubido}}]{gonzalez2022low}%
  \BibitemOpen
  \bibfield  {author} {\bibinfo {author} {\bibfnamefont {J.}~\bibnamefont {Gonz{\'a}lez}}, \bibinfo {author} {\bibfnamefont {D.}~\bibnamefont {Mateos}}, \bibinfo {author} {\bibfnamefont {M.}~\bibnamefont {Cavelli}}, \bibinfo {author} {\bibfnamefont {A.}~\bibnamefont {Mondino}}, \bibinfo {author} {\bibfnamefont {C.}~\bibnamefont {Pascovich}}, \bibinfo {author} {\bibfnamefont {P.}~\bibnamefont {Torterolo}},\ and\ \bibinfo {author} {\bibfnamefont {N.}~\bibnamefont {Rubido}},\ }\href@noop {} {\bibfield  {journal} {\bibinfo  {journal} {Neuroscience}\ }\textbf {\bibinfo {volume} {494}},\ \bibinfo {pages} {1} (\bibinfo {year} {2022})}\BibitemShut {NoStop}%
\bibitem [{\citenamefont {Gonz{\'a}lez}\ \emph {et~al.}(2023)\citenamefont {Gonz{\'a}lez}, \citenamefont {Cavelli}, \citenamefont {Tort}, \citenamefont {Torterolo},\ and\ \citenamefont {Rubido}}]{gonzalez2023sleep}%
  \BibitemOpen
  \bibfield  {author} {\bibinfo {author} {\bibfnamefont {J.}~\bibnamefont {Gonz{\'a}lez}}, \bibinfo {author} {\bibfnamefont {M.}~\bibnamefont {Cavelli}}, \bibinfo {author} {\bibfnamefont {A.~B.}\ \bibnamefont {Tort}}, \bibinfo {author} {\bibfnamefont {P.}~\bibnamefont {Torterolo}},\ and\ \bibinfo {author} {\bibfnamefont {N.}~\bibnamefont {Rubido}},\ }\href@noop {} {\bibfield  {journal} {\bibinfo  {journal} {Plos one}\ }\textbf {\bibinfo {volume} {18}},\ \bibinfo {pages} {e0290146} (\bibinfo {year} {2023})}\BibitemShut {NoStop}%
\bibitem [{\citenamefont {Bandt}(2023)}]{bandt2023statistics}%
  \BibitemOpen
  \bibfield  {author} {\bibinfo {author} {\bibfnamefont {C.}~\bibnamefont {Bandt}},\ }\href@noop {} {\bibfield  {journal} {\bibinfo  {journal} {Chaos: An Interdisciplinary Journal of Nonlinear Science}\ }\textbf {\bibinfo {volume} {33}},\ \bibinfo {pages} {033124} (\bibinfo {year} {2023})}\BibitemShut {NoStop}%
\bibitem [{\citenamefont {Zunino}(2024)}]{zunino2024revisiting}%
  \BibitemOpen
  \bibfield  {author} {\bibinfo {author} {\bibfnamefont {L.}~\bibnamefont {Zunino}},\ }\href@noop {} {\bibfield  {journal} {\bibinfo  {journal} {Entropy}\ }\textbf {\bibinfo {volume} {26}},\ \bibinfo {pages} {432} (\bibinfo {year} {2024})}\BibitemShut {NoStop}%
\bibitem [{\citenamefont {Boaretto}\ \emph {et~al.}(2023)\citenamefont {Boaretto}, \citenamefont {Budzinski}, \citenamefont {Rossi}, \citenamefont {Masoller},\ and\ \citenamefont {Macau}}]{boaretto2023spatial}%
  \BibitemOpen
  \bibfield  {author} {\bibinfo {author} {\bibfnamefont {B.~R.}\ \bibnamefont {Boaretto}}, \bibinfo {author} {\bibfnamefont {R.~C.}\ \bibnamefont {Budzinski}}, \bibinfo {author} {\bibfnamefont {K.~L.}\ \bibnamefont {Rossi}}, \bibinfo {author} {\bibfnamefont {C.}~\bibnamefont {Masoller}},\ and\ \bibinfo {author} {\bibfnamefont {E.~E.}\ \bibnamefont {Macau}},\ }\href@noop {} {\bibfield  {journal} {\bibinfo  {journal} {Chaos, Solitons \& Fractals}\ }\textbf {\bibinfo {volume} {171}},\ \bibinfo {pages} {113453} (\bibinfo {year} {2023})}\BibitemShut {NoStop}%
\bibitem [{\citenamefont {Gancio}\ \emph {et~al.}(2024{\natexlab{a}})\citenamefont {Gancio}, \citenamefont {Masoller},\ and\ \citenamefont {Tirabassi}}]{gancio2024permutation}%
  \BibitemOpen
  \bibfield  {author} {\bibinfo {author} {\bibfnamefont {J.}~\bibnamefont {Gancio}}, \bibinfo {author} {\bibfnamefont {C.}~\bibnamefont {Masoller}},\ and\ \bibinfo {author} {\bibfnamefont {G.}~\bibnamefont {Tirabassi}},\ }\href@noop {} {\bibfield  {journal} {\bibinfo  {journal} {Chaos: An Interdisciplinary Journal of Nonlinear Science}\ }\textbf {\bibinfo {volume} {34}},\ \bibinfo {pages} {043130} (\bibinfo {year} {2024}{\natexlab{a}})}\BibitemShut {NoStop}%
\bibitem [{\citenamefont {Rubido}\ \emph {et~al.}(2011)\citenamefont {Rubido}, \citenamefont {Tiana-Alsina}, \citenamefont {Torrent}, \citenamefont {Garcia-Ojalvo},\ and\ \citenamefont {Masoller}}]{rubido2011language}%
  \BibitemOpen
  \bibfield  {author} {\bibinfo {author} {\bibfnamefont {N.}~\bibnamefont {Rubido}}, \bibinfo {author} {\bibfnamefont {J.}~\bibnamefont {Tiana-Alsina}}, \bibinfo {author} {\bibfnamefont {M.}~\bibnamefont {Torrent}}, \bibinfo {author} {\bibfnamefont {J.}~\bibnamefont {Garcia-Ojalvo}},\ and\ \bibinfo {author} {\bibfnamefont {C.}~\bibnamefont {Masoller}},\ }\href@noop {} {\bibfield  {journal} {\bibinfo  {journal} {Physical Review E—Statistical, Nonlinear, and Soft Matter Physics}\ }\textbf {\bibinfo {volume} {84}},\ \bibinfo {pages} {026202} (\bibinfo {year} {2011})}\BibitemShut {NoStop}%
\bibitem [{\citenamefont {Soriano}\ \emph {et~al.}(2011)\citenamefont {Soriano}, \citenamefont {Zunino}, \citenamefont {Rosso}, \citenamefont {Fischer},\ and\ \citenamefont {Mirasso}}]{soriano2011time}%
  \BibitemOpen
  \bibfield  {author} {\bibinfo {author} {\bibfnamefont {M.~C.}\ \bibnamefont {Soriano}}, \bibinfo {author} {\bibfnamefont {L.}~\bibnamefont {Zunino}}, \bibinfo {author} {\bibfnamefont {O.~A.}\ \bibnamefont {Rosso}}, \bibinfo {author} {\bibfnamefont {I.}~\bibnamefont {Fischer}},\ and\ \bibinfo {author} {\bibfnamefont {C.~R.}\ \bibnamefont {Mirasso}},\ }\href@noop {} {\bibfield  {journal} {\bibinfo  {journal} {IEEE Journal of Quantum Electronics}\ }\textbf {\bibinfo {volume} {47}},\ \bibinfo {pages} {252} (\bibinfo {year} {2011})}\BibitemShut {NoStop}%
\bibitem [{\citenamefont {Aragoneses}\ \emph {et~al.}(2013)\citenamefont {Aragoneses}, \citenamefont {Rubido}, \citenamefont {Tiana-Alsina}, \citenamefont {Torrent},\ and\ \citenamefont {Masoller}}]{aragoneses2013distinguishing}%
  \BibitemOpen
  \bibfield  {author} {\bibinfo {author} {\bibfnamefont {A.}~\bibnamefont {Aragoneses}}, \bibinfo {author} {\bibfnamefont {N.}~\bibnamefont {Rubido}}, \bibinfo {author} {\bibfnamefont {J.}~\bibnamefont {Tiana-Alsina}}, \bibinfo {author} {\bibfnamefont {M.}~\bibnamefont {Torrent}},\ and\ \bibinfo {author} {\bibfnamefont {C.}~\bibnamefont {Masoller}},\ }\href@noop {} {\bibfield  {journal} {\bibinfo  {journal} {Scientific reports}\ }\textbf {\bibinfo {volume} {3}},\ \bibinfo {pages} {1778} (\bibinfo {year} {2013})}\BibitemShut {NoStop}%
\bibitem [{\citenamefont {Aragoneses}\ \emph {et~al.}(2014)\citenamefont {Aragoneses}, \citenamefont {Perrone}, \citenamefont {Sorrentino}, \citenamefont {Torrent},\ and\ \citenamefont {Masoller}}]{aragoneses2014unveiling}%
  \BibitemOpen
  \bibfield  {author} {\bibinfo {author} {\bibfnamefont {A.}~\bibnamefont {Aragoneses}}, \bibinfo {author} {\bibfnamefont {S.}~\bibnamefont {Perrone}}, \bibinfo {author} {\bibfnamefont {T.}~\bibnamefont {Sorrentino}}, \bibinfo {author} {\bibfnamefont {M.}~\bibnamefont {Torrent}},\ and\ \bibinfo {author} {\bibfnamefont {C.}~\bibnamefont {Masoller}},\ }\href@noop {} {\bibfield  {journal} {\bibinfo  {journal} {Scientific reports}\ }\textbf {\bibinfo {volume} {4}},\ \bibinfo {pages} {4696} (\bibinfo {year} {2014})}\BibitemShut {NoStop}%
\bibitem [{\citenamefont {Aragoneses}\ \emph {et~al.}(2016)\citenamefont {Aragoneses}, \citenamefont {Carpi}, \citenamefont {Tarasov}, \citenamefont {Churkin}, \citenamefont {Torrent}, \citenamefont {Masoller},\ and\ \citenamefont {Turitsyn}}]{aragoneses2016unveiling}%
  \BibitemOpen
  \bibfield  {author} {\bibinfo {author} {\bibfnamefont {A.}~\bibnamefont {Aragoneses}}, \bibinfo {author} {\bibfnamefont {L.}~\bibnamefont {Carpi}}, \bibinfo {author} {\bibfnamefont {N.}~\bibnamefont {Tarasov}}, \bibinfo {author} {\bibfnamefont {D.}~\bibnamefont {Churkin}}, \bibinfo {author} {\bibfnamefont {M.}~\bibnamefont {Torrent}}, \bibinfo {author} {\bibfnamefont {C.}~\bibnamefont {Masoller}},\ and\ \bibinfo {author} {\bibfnamefont {S.}~\bibnamefont {Turitsyn}},\ }\href@noop {} {\bibfield  {journal} {\bibinfo  {journal} {Physical review letters}\ }\textbf {\bibinfo {volume} {116}},\ \bibinfo {pages} {033902} (\bibinfo {year} {2016})}\BibitemShut {NoStop}%
\bibitem [{\citenamefont {Tirabassi}\ \emph {et~al.}(2023)\citenamefont {Tirabassi}, \citenamefont {Duque-Gijon}, \citenamefont {Tiana-Alsina},\ and\ \citenamefont {Masoller}}]{tirabassi2023permutation}%
  \BibitemOpen
  \bibfield  {author} {\bibinfo {author} {\bibfnamefont {G.}~\bibnamefont {Tirabassi}}, \bibinfo {author} {\bibfnamefont {M.}~\bibnamefont {Duque-Gijon}}, \bibinfo {author} {\bibfnamefont {J.}~\bibnamefont {Tiana-Alsina}},\ and\ \bibinfo {author} {\bibfnamefont {C.}~\bibnamefont {Masoller}},\ }\href@noop {} {\bibfield  {journal} {\bibinfo  {journal} {APL Photonics}\ }\textbf {\bibinfo {volume} {8}},\ \bibinfo {pages} {126112} (\bibinfo {year} {2023})}\BibitemShut {NoStop}%
\bibitem [{\citenamefont {Boaretto}\ \emph {et~al.}(2024)\citenamefont {Boaretto}, \citenamefont {Macau},\ and\ \citenamefont {Masoller}}]{boaretto2024characterizing}%
  \BibitemOpen
  \bibfield  {author} {\bibinfo {author} {\bibfnamefont {B.~R.}\ \bibnamefont {Boaretto}}, \bibinfo {author} {\bibfnamefont {E.~E.}\ \bibnamefont {Macau}},\ and\ \bibinfo {author} {\bibfnamefont {C.}~\bibnamefont {Masoller}},\ }\href@noop {} {\bibfield  {journal} {\bibinfo  {journal} {Chaos: An Interdisciplinary Journal of Nonlinear Science}\ }\textbf {\bibinfo {volume} {34}},\ \bibinfo {pages} {043108} (\bibinfo {year} {2024})}\BibitemShut {NoStop}%
\bibitem [{\citenamefont {Zunino}\ \emph {et~al.}(2024)\citenamefont {Zunino}, \citenamefont {Porte},\ and\ \citenamefont {Soriano}}]{zunino2024identifying}%
  \BibitemOpen
  \bibfield  {author} {\bibinfo {author} {\bibfnamefont {L.}~\bibnamefont {Zunino}}, \bibinfo {author} {\bibfnamefont {X.}~\bibnamefont {Porte}},\ and\ \bibinfo {author} {\bibfnamefont {M.~C.}\ \bibnamefont {Soriano}},\ }\href@noop {} {\bibfield  {journal} {\bibinfo  {journal} {Entropy}\ }\textbf {\bibinfo {volume} {26}},\ \bibinfo {pages} {1016} (\bibinfo {year} {2024})}\BibitemShut {NoStop}%
\bibitem [{\citenamefont {Barreiro}\ \emph {et~al.}(2011)\citenamefont {Barreiro}, \citenamefont {Marti},\ and\ \citenamefont {Masoller}}]{barreiro2011inferring}%
  \BibitemOpen
  \bibfield  {author} {\bibinfo {author} {\bibfnamefont {M.}~\bibnamefont {Barreiro}}, \bibinfo {author} {\bibfnamefont {A.~C.}\ \bibnamefont {Marti}},\ and\ \bibinfo {author} {\bibfnamefont {C.}~\bibnamefont {Masoller}},\ }\href@noop {} {\bibfield  {journal} {\bibinfo  {journal} {Chaos: An Interdisciplinary Journal of Nonlinear Science}\ }\textbf {\bibinfo {volume} {21}},\ \bibinfo {pages} {013101} (\bibinfo {year} {2011})}\BibitemShut {NoStop}%
\bibitem [{\citenamefont {Deza}\ \emph {et~al.}(2013)\citenamefont {Deza}, \citenamefont {Barreiro},\ and\ \citenamefont {Masoller}}]{deza2013inferring}%
  \BibitemOpen
  \bibfield  {author} {\bibinfo {author} {\bibfnamefont {J.~I.}\ \bibnamefont {Deza}}, \bibinfo {author} {\bibfnamefont {M.}~\bibnamefont {Barreiro}},\ and\ \bibinfo {author} {\bibfnamefont {C.}~\bibnamefont {Masoller}},\ }\href@noop {} {\bibfield  {journal} {\bibinfo  {journal} {The European Physical Journal Special Topics}\ }\textbf {\bibinfo {volume} {222}},\ \bibinfo {pages} {511} (\bibinfo {year} {2013})}\BibitemShut {NoStop}%
\bibitem [{\citenamefont {Tupikina}\ \emph {et~al.}(2014)\citenamefont {Tupikina}, \citenamefont {Rehfeld}, \citenamefont {Molkenthin}, \citenamefont {Stolbova}, \citenamefont {Marwan},\ and\ \citenamefont {Kurths}}]{tupikina2014characterizing}%
  \BibitemOpen
  \bibfield  {author} {\bibinfo {author} {\bibfnamefont {L.}~\bibnamefont {Tupikina}}, \bibinfo {author} {\bibfnamefont {K.}~\bibnamefont {Rehfeld}}, \bibinfo {author} {\bibfnamefont {N.}~\bibnamefont {Molkenthin}}, \bibinfo {author} {\bibfnamefont {V.}~\bibnamefont {Stolbova}}, \bibinfo {author} {\bibfnamefont {N.}~\bibnamefont {Marwan}},\ and\ \bibinfo {author} {\bibfnamefont {J.}~\bibnamefont {Kurths}},\ }\href@noop {} {\bibfield  {journal} {\bibinfo  {journal} {Nonlinear Processes in Geophysics}\ }\textbf {\bibinfo {volume} {21}},\ \bibinfo {pages} {705} (\bibinfo {year} {2014})}\BibitemShut {NoStop}%
\bibitem [{\citenamefont {Deza}\ \emph {et~al.}(2018)\citenamefont {Deza}, \citenamefont {Tirabassi}, \citenamefont {Barreiro},\ and\ \citenamefont {Masoller}}]{deza2018large}%
  \BibitemOpen
  \bibfield  {author} {\bibinfo {author} {\bibfnamefont {J.}~\bibnamefont {Deza}}, \bibinfo {author} {\bibfnamefont {G.}~\bibnamefont {Tirabassi}}, \bibinfo {author} {\bibfnamefont {M.}~\bibnamefont {Barreiro}},\ and\ \bibinfo {author} {\bibfnamefont {C.}~\bibnamefont {Masoller}},\ }\href@noop {} {\bibfield  {journal} {\bibinfo  {journal} {Advances in Nonlinear Geosciences}\ ,\ \bibinfo {pages} {87}} (\bibinfo {year} {2018})}\BibitemShut {NoStop}%
\bibitem [{\citenamefont {Dijkstra}\ \emph {et~al.}(2019)\citenamefont {Dijkstra}, \citenamefont {Hern{\'a}ndez-Garc{\'\i}a}, \citenamefont {Masoller},\ and\ \citenamefont {Barreiro}}]{dijkstra2019networks}%
  \BibitemOpen
  \bibfield  {author} {\bibinfo {author} {\bibfnamefont {H.~A.}\ \bibnamefont {Dijkstra}}, \bibinfo {author} {\bibfnamefont {E.}~\bibnamefont {Hern{\'a}ndez-Garc{\'\i}a}}, \bibinfo {author} {\bibfnamefont {C.}~\bibnamefont {Masoller}},\ and\ \bibinfo {author} {\bibfnamefont {M.}~\bibnamefont {Barreiro}},\ }\href@noop {} {\emph {\bibinfo {title} {Networks in climate}}}\ (\bibinfo  {publisher} {Cambridge University Press},\ \bibinfo {year} {2019})\BibitemShut {NoStop}%
\bibitem [{\citenamefont {Wu}\ \emph {et~al.}(2020)\citenamefont {Wu}, \citenamefont {Zou}, \citenamefont {Alves}, \citenamefont {Macau}, \citenamefont {Sampaio},\ and\ \citenamefont {Marengo}}]{wu2020uncovering}%
  \BibitemOpen
  \bibfield  {author} {\bibinfo {author} {\bibfnamefont {H.}~\bibnamefont {Wu}}, \bibinfo {author} {\bibfnamefont {Y.}~\bibnamefont {Zou}}, \bibinfo {author} {\bibfnamefont {L.~M.}\ \bibnamefont {Alves}}, \bibinfo {author} {\bibfnamefont {E.~E.}\ \bibnamefont {Macau}}, \bibinfo {author} {\bibfnamefont {G.}~\bibnamefont {Sampaio}},\ and\ \bibinfo {author} {\bibfnamefont {J.~A.}\ \bibnamefont {Marengo}},\ }\href@noop {} {\bibfield  {journal} {\bibinfo  {journal} {Chaos: An Interdisciplinary Journal of Nonlinear Science}\ }\textbf {\bibinfo {volume} {30}},\ \bibinfo {pages} {053104} (\bibinfo {year} {2020})}\BibitemShut {NoStop}%
\bibitem [{\citenamefont {Ruiz-Aguilar}\ \emph {et~al.}(2021)\citenamefont {Ruiz-Aguilar}, \citenamefont {Turias}, \citenamefont {Gonz{\'a}lez-Enrique}, \citenamefont {Urda},\ and\ \citenamefont {Elizondo}}]{ruiz2021permutation}%
  \BibitemOpen
  \bibfield  {author} {\bibinfo {author} {\bibfnamefont {J.~J.}\ \bibnamefont {Ruiz-Aguilar}}, \bibinfo {author} {\bibfnamefont {I.}~\bibnamefont {Turias}}, \bibinfo {author} {\bibfnamefont {J.}~\bibnamefont {Gonz{\'a}lez-Enrique}}, \bibinfo {author} {\bibfnamefont {D.}~\bibnamefont {Urda}},\ and\ \bibinfo {author} {\bibfnamefont {D.}~\bibnamefont {Elizondo}},\ }\href@noop {} {\bibfield  {journal} {\bibinfo  {journal} {Neural Computing and Applications}\ }\textbf {\bibinfo {volume} {33}},\ \bibinfo {pages} {2369} (\bibinfo {year} {2021})}\BibitemShut {NoStop}%
\bibitem [{\citenamefont {Gancio}\ \emph {et~al.}(2024{\natexlab{b}})\citenamefont {Gancio}, \citenamefont {Tirabassi}, \citenamefont {Masoller},\ and\ \citenamefont {Barreiro}}]{gancio2024analysis}%
  \BibitemOpen
  \bibfield  {author} {\bibinfo {author} {\bibfnamefont {J.}~\bibnamefont {Gancio}}, \bibinfo {author} {\bibfnamefont {G.}~\bibnamefont {Tirabassi}}, \bibinfo {author} {\bibfnamefont {C.}~\bibnamefont {Masoller}},\ and\ \bibinfo {author} {\bibfnamefont {M.}~\bibnamefont {Barreiro}},\ }\href@noop {} {\bibfield  {journal} {\bibinfo  {journal} {Earth System Dynamics Discussions}\ }\textbf {\bibinfo {volume} {2024}},\ \bibinfo {pages} {1} (\bibinfo {year} {2024}{\natexlab{b}})}\BibitemShut {NoStop}%
\bibitem [{\citenamefont {Zanin}(2008)}]{zanin2008forbidden}%
  \BibitemOpen
  \bibfield  {author} {\bibinfo {author} {\bibfnamefont {M.}~\bibnamefont {Zanin}},\ }\href@noop {} {\bibfield  {journal} {\bibinfo  {journal} {Chaos: An Interdisciplinary Journal of Nonlinear Science}\ }\textbf {\bibinfo {volume} {18}},\ \bibinfo {pages} {013119} (\bibinfo {year} {2008})}\BibitemShut {NoStop}%
\bibitem [{\citenamefont {Zunino}\ \emph {et~al.}(2009)\citenamefont {Zunino}, \citenamefont {Zanin}, \citenamefont {Tabak}, \citenamefont {P{\'e}rez},\ and\ \citenamefont {Rosso}}]{zunino2009forbidden}%
  \BibitemOpen
  \bibfield  {author} {\bibinfo {author} {\bibfnamefont {L.}~\bibnamefont {Zunino}}, \bibinfo {author} {\bibfnamefont {M.}~\bibnamefont {Zanin}}, \bibinfo {author} {\bibfnamefont {B.~M.}\ \bibnamefont {Tabak}}, \bibinfo {author} {\bibfnamefont {D.~G.}\ \bibnamefont {P{\'e}rez}},\ and\ \bibinfo {author} {\bibfnamefont {O.~A.}\ \bibnamefont {Rosso}},\ }\href@noop {} {\bibfield  {journal} {\bibinfo  {journal} {Physica A: Statistical Mechanics and its Applications}\ }\textbf {\bibinfo {volume} {388}},\ \bibinfo {pages} {2854} (\bibinfo {year} {2009})}\BibitemShut {NoStop}%
\bibitem [{\citenamefont {Zhao}\ \emph {et~al.}(2013)\citenamefont {Zhao}, \citenamefont {Shang},\ and\ \citenamefont {Wang}}]{zhao2013measuring}%
  \BibitemOpen
  \bibfield  {author} {\bibinfo {author} {\bibfnamefont {X.}~\bibnamefont {Zhao}}, \bibinfo {author} {\bibfnamefont {P.}~\bibnamefont {Shang}},\ and\ \bibinfo {author} {\bibfnamefont {J.}~\bibnamefont {Wang}},\ }\href@noop {} {\bibfield  {journal} {\bibinfo  {journal} {Physical Review E—Statistical, Nonlinear, and Soft Matter Physics}\ }\textbf {\bibinfo {volume} {87}},\ \bibinfo {pages} {022805} (\bibinfo {year} {2013})}\BibitemShut {NoStop}%
\bibitem [{\citenamefont {Yin}\ and\ \citenamefont {Shang}(2014)}]{yin2014weighted}%
  \BibitemOpen
  \bibfield  {author} {\bibinfo {author} {\bibfnamefont {Y.}~\bibnamefont {Yin}}\ and\ \bibinfo {author} {\bibfnamefont {P.}~\bibnamefont {Shang}},\ }\href@noop {} {\bibfield  {journal} {\bibinfo  {journal} {Nonlinear Dynamics}\ }\textbf {\bibinfo {volume} {78}},\ \bibinfo {pages} {2921} (\bibinfo {year} {2014})}\BibitemShut {NoStop}%
\bibitem [{\citenamefont {Stosic}\ \emph {et~al.}(2019)\citenamefont {Stosic}, \citenamefont {Stosic}, \citenamefont {Ludermir},\ and\ \citenamefont {Stosic}}]{stosic2019exploring}%
  \BibitemOpen
  \bibfield  {author} {\bibinfo {author} {\bibfnamefont {D.}~\bibnamefont {Stosic}}, \bibinfo {author} {\bibfnamefont {D.}~\bibnamefont {Stosic}}, \bibinfo {author} {\bibfnamefont {T.~B.}\ \bibnamefont {Ludermir}},\ and\ \bibinfo {author} {\bibfnamefont {T.}~\bibnamefont {Stosic}},\ }\href@noop {} {\bibfield  {journal} {\bibinfo  {journal} {Physica A: Statistical Mechanics and its Applications}\ }\textbf {\bibinfo {volume} {525}},\ \bibinfo {pages} {548} (\bibinfo {year} {2019})}\BibitemShut {NoStop}%
\bibitem [{\citenamefont {Henry}\ and\ \citenamefont {Judge}(2019)}]{henry2019permutation}%
  \BibitemOpen
  \bibfield  {author} {\bibinfo {author} {\bibfnamefont {M.}~\bibnamefont {Henry}}\ and\ \bibinfo {author} {\bibfnamefont {G.}~\bibnamefont {Judge}},\ }\href@noop {} {\bibfield  {journal} {\bibinfo  {journal} {Econometrics}\ }\textbf {\bibinfo {volume} {7}},\ \bibinfo {pages} {10} (\bibinfo {year} {2019})}\BibitemShut {NoStop}%
\bibitem [{\citenamefont {Bandt}(2020)}]{bandt2020order}%
  \BibitemOpen
  \bibfield  {author} {\bibinfo {author} {\bibfnamefont {C.}~\bibnamefont {Bandt}},\ }\href@noop {} {\bibfield  {journal} {\bibinfo  {journal} {Statistical Papers}\ }\textbf {\bibinfo {volume} {61}},\ \bibinfo {pages} {1565} (\bibinfo {year} {2020})}\BibitemShut {NoStop}%
\bibitem [{\citenamefont {Kozak}\ \emph {et~al.}(2020)\citenamefont {Kozak}, \citenamefont {Kania},\ and\ \citenamefont {Juszczuk}}]{kozak2020permutation}%
  \BibitemOpen
  \bibfield  {author} {\bibinfo {author} {\bibfnamefont {J.}~\bibnamefont {Kozak}}, \bibinfo {author} {\bibfnamefont {K.}~\bibnamefont {Kania}},\ and\ \bibinfo {author} {\bibfnamefont {P.}~\bibnamefont {Juszczuk}},\ }\href@noop {} {\bibfield  {journal} {\bibinfo  {journal} {Entropy}\ }\textbf {\bibinfo {volume} {22}},\ \bibinfo {pages} {330} (\bibinfo {year} {2020})}\BibitemShut {NoStop}%
\bibitem [{\citenamefont {R{\'e}nyi}(1961)}]{renyi1961measures}%
  \BibitemOpen
  \bibfield  {author} {\bibinfo {author} {\bibfnamefont {A.}~\bibnamefont {R{\'e}nyi}},\ }in\ \href@noop {} {\emph {\bibinfo {booktitle} {Proceedings of the fourth Berkeley symposium on mathematical statistics and probability, volume 1: contributions to the theory of statistics}}},\ Vol.~\bibinfo {volume} {4}\ (\bibinfo {organization} {University of California Press},\ \bibinfo {year} {1961})\ pp.\ \bibinfo {pages} {547--562}\BibitemShut {NoStop}%
\bibitem [{\citenamefont {May}(1976)}]{may1976simple}%
  \BibitemOpen
  \bibfield  {author} {\bibinfo {author} {\bibfnamefont {R.~M.}\ \bibnamefont {May}},\ }\href@noop {} {\bibfield  {journal} {\bibinfo  {journal} {Nature}\ }\textbf {\bibinfo {volume} {261}},\ \bibinfo {pages} {459} (\bibinfo {year} {1976})}\BibitemShut {NoStop}%
\bibitem [{\citenamefont {Kaneko}(1990)}]{kaneko1990clustering}%
  \BibitemOpen
  \bibfield  {author} {\bibinfo {author} {\bibfnamefont {K.}~\bibnamefont {Kaneko}},\ }\href@noop {} {\bibfield  {journal} {\bibinfo  {journal} {Physica D: Nonlinear Phenomena}\ }\textbf {\bibinfo {volume} {41}},\ \bibinfo {pages} {137} (\bibinfo {year} {1990})}\BibitemShut {NoStop}%
\bibitem [{\citenamefont {L’Her}\ \emph {et~al.}(2016)\citenamefont {L’Her}, \citenamefont {Amil}, \citenamefont {Rubido}, \citenamefont {Marti},\ and\ \citenamefont {Cabeza}}]{l2016electronically}%
  \BibitemOpen
  \bibfield  {author} {\bibinfo {author} {\bibfnamefont {A.}~\bibnamefont {L’Her}}, \bibinfo {author} {\bibfnamefont {P.}~\bibnamefont {Amil}}, \bibinfo {author} {\bibfnamefont {N.}~\bibnamefont {Rubido}}, \bibinfo {author} {\bibfnamefont {A.~C.}\ \bibnamefont {Marti}},\ and\ \bibinfo {author} {\bibfnamefont {C.}~\bibnamefont {Cabeza}},\ }\href@noop {} {\bibfield  {journal} {\bibinfo  {journal} {The European Physical Journal B}\ }\textbf {\bibinfo {volume} {89}},\ \bibinfo {pages} {1} (\bibinfo {year} {2016})}\BibitemShut {NoStop}%
\bibitem [{\citenamefont {H{\'e}non}(2004)}]{henon2004two}%
  \BibitemOpen
  \bibfield  {author} {\bibinfo {author} {\bibfnamefont {M.}~\bibnamefont {H{\'e}non}},\ }\href@noop {} {\bibfield  {journal} {\bibinfo  {journal} {The theory of chaotic attractors}\ ,\ \bibinfo {pages} {94}} (\bibinfo {year} {2004})}\BibitemShut {NoStop}%
\bibitem [{\citenamefont {Politi}(2017)}]{politi2017quantifying}%
  \BibitemOpen
  \bibfield  {author} {\bibinfo {author} {\bibfnamefont {A.}~\bibnamefont {Politi}},\ }\href@noop {} {\bibfield  {journal} {\bibinfo  {journal} {Physical review letters}\ }\textbf {\bibinfo {volume} {118}},\ \bibinfo {pages} {144101} (\bibinfo {year} {2017})}\BibitemShut {NoStop}%
\bibitem [{\citenamefont {Watt}\ and\ \citenamefont {Politi}(2019)}]{watt2019permutation}%
  \BibitemOpen
  \bibfield  {author} {\bibinfo {author} {\bibfnamefont {S.~J.}\ \bibnamefont {Watt}}\ and\ \bibinfo {author} {\bibfnamefont {A.}~\bibnamefont {Politi}},\ }\href@noop {} {\bibfield  {journal} {\bibinfo  {journal} {Chaos, Solitons \& Fractals}\ }\textbf {\bibinfo {volume} {120}},\ \bibinfo {pages} {95} (\bibinfo {year} {2019})}\BibitemShut {NoStop}%
\bibitem [{\citenamefont {R{\'e}nyi}(1959)}]{renyi1959dimension}%
  \BibitemOpen
  \bibfield  {author} {\bibinfo {author} {\bibfnamefont {A.}~\bibnamefont {R{\'e}nyi}},\ }\href@noop {} {\bibfield  {journal} {\bibinfo  {journal} {Acta Mathematica Academiae Scientiarum Hungarica}\ }\textbf {\bibinfo {volume} {10}},\ \bibinfo {pages} {193} (\bibinfo {year} {1959})}\BibitemShut {NoStop}%
\bibitem [{\citenamefont {Kolmogorov}(1959)}]{kolmogorov1959entropy}%
  \BibitemOpen
  \bibfield  {author} {\bibinfo {author} {\bibfnamefont {A.~N.}\ \bibnamefont {Kolmogorov}},\ }in\ \href@noop {} {\emph {\bibinfo {booktitle} {Dokl. Akad. Nauk SSSR}}},\ Vol.\ \bibinfo {volume} {124}\ (\bibinfo {year} {1959})\ pp.\ \bibinfo {pages} {754--755}\BibitemShut {NoStop}%
\bibitem [{\citenamefont {Sinai}(1959)}]{sinai1959notion}%
  \BibitemOpen
  \bibfield  {author} {\bibinfo {author} {\bibfnamefont {Y.~G.}\ \bibnamefont {Sinai}},\ }in\ \href@noop {} {\emph {\bibinfo {booktitle} {Doklady of Russian Academy of Sciences}}},\ Vol.\ \bibinfo {volume} {124}\ (\bibinfo {year} {1959})\ pp.\ \bibinfo {pages} {768--771}\BibitemShut {NoStop}%
\bibitem [{\citenamefont {Bandt}\ \emph {et~al.}(2002)\citenamefont {Bandt}, \citenamefont {Keller},\ and\ \citenamefont {Pompe}}]{bandt2002entropy}%
  \BibitemOpen
  \bibfield  {author} {\bibinfo {author} {\bibfnamefont {C.}~\bibnamefont {Bandt}}, \bibinfo {author} {\bibfnamefont {G.}~\bibnamefont {Keller}},\ and\ \bibinfo {author} {\bibfnamefont {B.}~\bibnamefont {Pompe}},\ }\href@noop {} {\bibfield  {journal} {\bibinfo  {journal} {Nonlinearity}\ }\textbf {\bibinfo {volume} {15}},\ \bibinfo {pages} {1595} (\bibinfo {year} {2002})}\BibitemShut {NoStop}%
\bibitem [{\citenamefont {Amig{\'o}}\ \emph {et~al.}(2005)\citenamefont {Amig{\'o}}, \citenamefont {Kennel},\ and\ \citenamefont {Kocarev}}]{amigo2005permutation}%
  \BibitemOpen
  \bibfield  {author} {\bibinfo {author} {\bibfnamefont {J.~M.}\ \bibnamefont {Amig{\'o}}}, \bibinfo {author} {\bibfnamefont {M.~B.}\ \bibnamefont {Kennel}},\ and\ \bibinfo {author} {\bibfnamefont {L.}~\bibnamefont {Kocarev}},\ }\href@noop {} {\bibfield  {journal} {\bibinfo  {journal} {Physica D: Nonlinear Phenomena}\ }\textbf {\bibinfo {volume} {210}},\ \bibinfo {pages} {77} (\bibinfo {year} {2005})}\BibitemShut {NoStop}%
\end{thebibliography}
\end{document}